\pdfoutput=1

\documentclass[11pt,a4paper]{article}
\usepackage{jheppub}
\usepackage{appendix}
\usepackage{graphicx}
\usepackage{caption}
\usepackage{subcaption}
\usepackage{footmisc}
\usepackage{amsmath}

\newcommand{\pa}{\partial}

\title{Gravitational waves from a holographic phase transition}

\author[a,b,c]{F\"eanor Reuben Ares,} 
\author[a,b,c]{Mark Hindmarsh,}
\author[d]{Carlos Hoyos,}
\author{and}
\author[b,c]{Niko Jokela}
\affiliation[a]{Department of Physics and Astronomy, University of Sussex,\\
 Brighton BN1 9QH, UK}
\affiliation[b]{Department of Physics and ${}^{c}$Helsinki Institute of Physics,\\
PL 64, 00014 University of Helsinki, Finland}
\affiliation[d]{Department of Physics \& Instituto de Ciencias y Tecnolog\'{\i}as Espaciales de Asturias (ICTEA), Universidad de Oviedo, 
c/. Federico Garc\'ia Lorca 18, ES-33007 Oviedo, Spain}
\emailAdd{F.R.Ares@sussex.ac.uk} 
\emailAdd{mark.hindmarsh@helsinki.fi} 
\emailAdd{hoyoscarlos@uniovi.es}
\emailAdd{niko.jokela@helsinki.fi}

\abstract{
We investigate first order phase transitions in a holographic setting of five-dimensional Einstein gravity coupled to a scalar field, constructing phase diagrams of the dual field theory at finite temperature. We scan over the two-dimensional parameter space of a simple bottom-up model and map out important quantities for the phase transition: the region where first order phase transitions take place; the latent heat, the transition strength parameter $\alpha$, and the stiffness.  We find 
that $\alpha$ is generically in the range $0.1$ to $0.3$, and is strongly correlated with the stiffness 
(the square of the sound speed in a barotropic fluid). 
Using the LISA Cosmology Working Group gravitational wave power spectrum model 
corrected for kinetic energy suppression at large $\alpha$ and non-conformal stiffness, 
we outline the observational prospects at the future space-based detectors LISA and TianQin.  
A TeV-scale hidden sector with a phase transition described by the model could be observable 
at both detectors.
}

\keywords{Cosmology, Fluid/Gravity Correspondence, Gravitational Waves, Phase Transitions} 

\preprint{$\begin{array}{rr}
	\text{HIP-2020-31/TH}\\\text{Sussex-94886}\end{array}$
}

\begin{document}
\maketitle
\flushbottom

\setcounter{page}{2}

\clearpage

\section{Introduction}

Spontaneous symmetry breaking of gauge theories is a fundamental ingredient of nature, which can manifest itself in the early Universe as a phase transition \cite{Kirzhnits:1976ts,Linde:1978px}. 
In particular, when temperatures were in the range $100$ $-$ $1000$ $\text{GeV}$, there may have been a phase transition associated with the breaking of the electroweak symmetry. 
If this was a first order transition, gravitational waves would have been produced through bubble nucleation, collision and counter-propagating sound waves (see e.g.~\cite{Hindmarsh:2020hop}). 
There is a strong possibility that they would be of the right frequency to be observed by a space-based gravitational wave detector such as 
LISA (Laser Interferometer Space Antenna) \cite{amaroseoane2017laser}. LISA will be sensitive to gravitational waves in the frequency range $1$ to $10$ mHz with characteristic strains of order $10^{-21}$, and hence to phase transitions 
occurring at around $10^{-12}$ seconds after the big bang (see e.g.~\cite{Caprini:2019egz}).

The standard model electroweak transition is known to be a crossover \cite{Kajantie, Laine, Laine_2013}, however, even minimal extensions may allow a first-order transition \cite{Carena_1996,Delepine_1996,Laine_1998,Huber_2001,Grojean_2005,Huber_2007,Profumo_2007,Barger_2008,Dorsch:2013wja,Damgaard_2016}. One class of extensions invokes strong dynamics just above the electroweak scale, which triggers electroweak symmetry-breaking, while addressing the hierarchy problem (see e.g. \cite{Sannino:2009za,Cacciapaglia:2020kgq}). Strongly-coupled field theories are notoriously difficult to study quantitatively. Holography is a technique for simplifying calculation by translating these complex strongly-coupled field theories into more tractable weakly-coupled gravitational theories \cite{CasalderreySolana:2011us,Ramallo:2013bua,Brambilla:2014jmp}.

Cosmological phase transitions in holographic models have been studied mostly in the context of Randall-Sundrum models \cite{Creminelli:2001th,Randall:2006py,Nardini:2007me,Konstandin:2010cd,Bunk:2017fic,Dillon:2017ctw,Megias:2018sxv,Baratella:2018pxi,Agashe:2019lhy,Fujikura:2019oyi,DelleRose:2019pgi,vonHarling:2019gme,Megias:2020vek,agashe2020phase}, where there is a first order phase transition between a black brane geometry and a horizonless geometry. From the field theory dual point of view this is interpreted as a confinement transition \cite{Witten:1998qj}. More recently in \cite{Bigazzi:2020phm}, this approach has been extended to the confinement transition in a model based on a string theory construction, the Sakai-Sugimoto model \cite{Sakai:2004cn}.  In addition \cite{Bigazzi:2020phm} also studied chiral phase transitions in the same model. The chiral transition is realised through probe branes in a fixed black brane background entering the horizon \cite{Mateos:2006nu,Aharony:2006da}. 

Gravitational wave production has also been studied in Randall-Sundrum models  \cite{Randall:2006py,Megias:2018sxv,Baratella:2018pxi,Agashe:2019lhy,DelleRose:2019pgi,vonHarling:2019gme,Megias:2020vek,agashe2020phase}, and recently in the context of the above-mentioned Sakai-Sugimoto model \cite{Bigazzi:2020avc}. 
It should be noted that all these studies are based on static configurations. Dynamical evolution of strongly coupled theories close to a phase transition has been studied in the context of applications to heavy ion collisions and condensed matter \cite{Gursoy:2016ggq,Critelli:2018osu,Attems:2018gou,Bantilan:2020pay}, including dynamical phase separation in three-dimensional \cite{Janik:2017ykj,Bellantuono:2019wbn,Li:2020ayr} and four-dimensional  \cite{Attems:2017ezz,Attems:2019yqn,Bea:2020ees} theories.

In this paper we study a phase transition in a simple bottom-up holographic model, calculating the equilibrium parameters which are most important in determining gravitational wave signals.  We scan over the two parameters of the model, showing that the transitions are generically ``intermediate'' in strength in the classification of Ref.~\cite{Cutting_2020}, meaning 
that the transition strength parameter at the critical temperature 
(the fraction of the energy available for conversion to kinetic energy and thereby gravitational wave production) is 
$\alpha = {\rm O}(10^{-1})$. Strong transitions ($\alpha = {\rm O}(1)$) are also possible with supercooling. 
We find a strong correlation between $\alpha$ and the stiffness at the critical temperature, 
meaning that the speed of sound can be quite different from $1/\sqrt{3}$. 

We then study the implications for gravitational wave production and observation, 
using the LISA Cosmology Working Group model \cite{Caprini_2020} as a starting point. 
We take into account recent work on kinetic energy conversion at
strong transitions  \cite{Cutting_2020} and when the stiffness is different from $1/3$ \cite{Giese_2020}, 
and include an improved treatment of the effect of the finite lifetime of the source \cite{guo2020phase}.

We find that the transitions in the holographic model are strong enough to be easily seen at LISA (and the similarly configured Taiji \cite{Ruan_2020}), and possibly even TianQin \cite{Ming_2020}, if the peak frequency is in the range of the maximum sensitivity.   The condition of observability constrains a combination of the transition temperature, the transition rate parameter, and the wall speed. 
 
The rest of this paper is organised as follows. In Section~\ref{sec:1pt} we review the putative first order phase transitions and their relation to properties of expanding bubbles in a cosmological context. In Section~\ref{sec:holo} we will describe the holographic model \cite{Bea:2018whf} and its black brane solutions. In Section~\ref{sec:thermo} we compute the thermodynamic quantities of interest from the holographic model. Equipped with the equation of state across the phase transition, in Section~\ref{sec:parameterscanning} we make a scan over the free parameters of the holographic model and find the regions of the parameter space in which a strongly first order phase transition exists, and the relevant thermodynamic parameters for gravitational wave production. In Section~\ref{sec:GW} we will determine if the signal as extracted from the holographic model is in the sensitivity window of future gravitational wave detectors. We conclude in Section~\ref{sec:discussion} with a discussion of our findings and some thoughts on future developments of our work. Appendices~\ref{app:holoreno} and \ref{app:numerics} contain details of the holographic renormalisation and the numerical procedures we implement, respectively, and Appendix \ref{app:omegagw} contains details of the power spectrum model, 
describing the modifications to that of Ref.~\cite{Caprini_2020} we have introduced.

\section{First order phase transitions}\label{sec:1pt}

First-order transitions from an `old' to a `new' phase proceed through the nucleation of bubbles in the old phase, with an order parameter jumping discontinuously at the transition temperature. Coleman \cite{Fate,Fate2} was the first to analyse how a metastable phase could decay through vacuum quantum fluctuations via bubbles nucleating containing a stable phase at zero-temperature in a cosmological setting. Later on Linde \cite{1977PhLB...70..306L, LINDE198137} generalised Coleman's work to bubbles nucleating at a non-zero temperature. Collision of these bubbles would be an extremely energetic process, leading to gravitational waves being produced in a possibly observable way \cite{Witten:1984rs}. Accurately estimating the power spectra of the signal is of great import as detection of cosmological gravitational waves would be strong evidence for physics beyond the Standard Model (see   
\cite{Caprini_2020} for a review).

\begin{figure}
  \centering
      \includegraphics[width=0.75\textwidth]{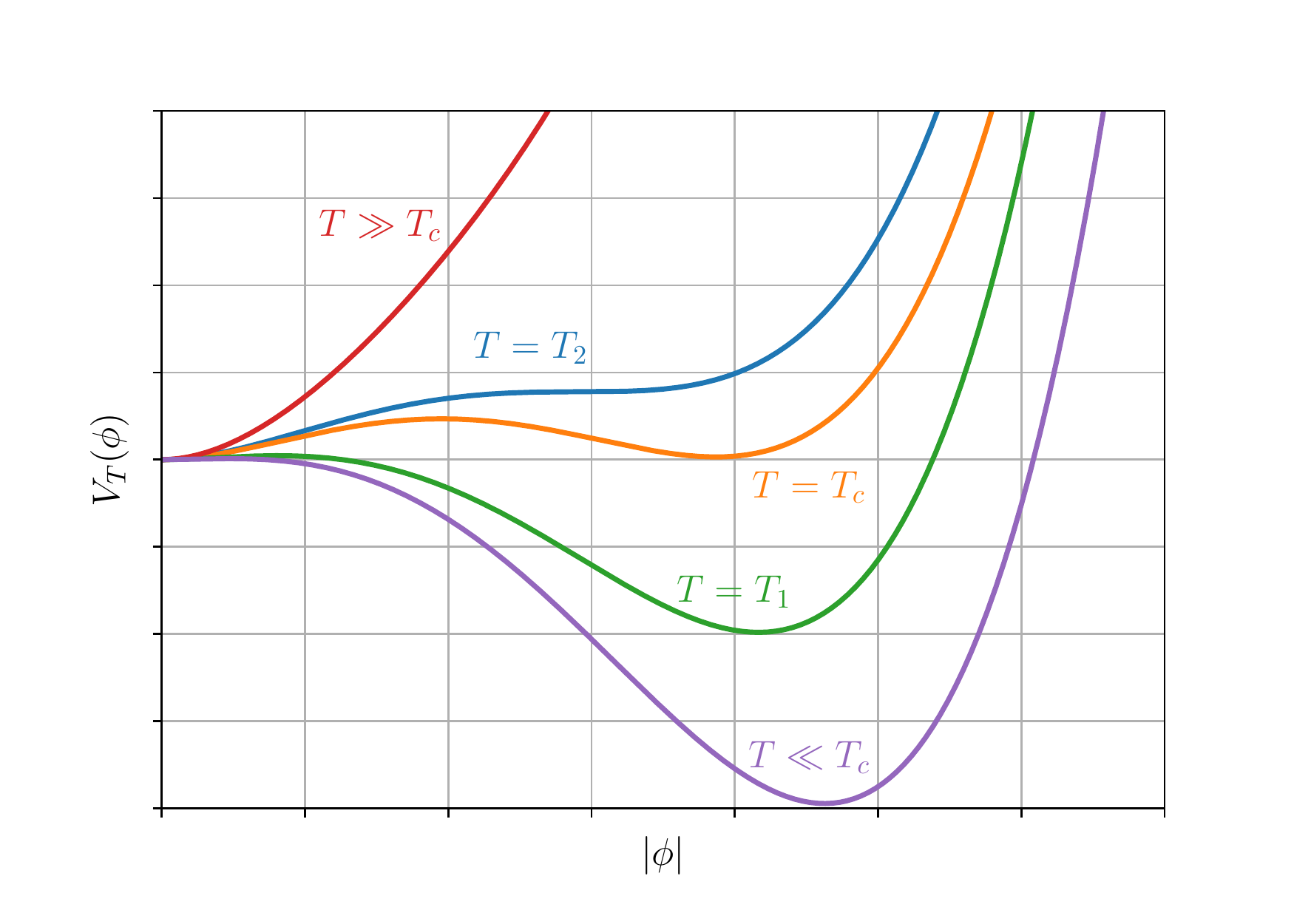}
  \caption{The thermal potential for varying temperatures. $T\gg T_c$ is the limit in which the potential is symmetric around $\phi = 0$, $T = T_2$ is the temperature below which a second minimum appears, $T_c$ is the critical temperature at which there are degenerate minima, and $T = T_1$ is where the first minimum at the origin disappears and the second minimum becomes the only equilibrium state. 
  }
  \label{fig:quartic}
\end{figure}
Fluctuations in the old phase trigger the nucleation of bubbles of the new phase. These bubbles would then collide and merge until the Universe would saturate with the new phase, at which time the phase transition would be complete.
The generation of bubbles and whether conditions are right for them to proliferate is described by four main parameters: transition strength $\alpha$, transition rate $\beta$, nucleation temperature $T_N$, and the wall speed $v_w$. 
Recently, it has also been pointed out that the sound speed, controlled by the stiffness $\pa p/\pa e$ is also  important \cite{Garny:2020rom,Giese_2020}.

The important temperatures in a phase transition are as follows. First, the critical temperature $T_c$, where the free energy of two competing phases first becomes equal, as shown in Fig.~\ref{fig:quartic}. 
Bubble nucleation takes place at a lower temperature $T_N<T_c$, 
where the phase transition actually takes place.  
Between these two temperatures the system is in a supercooled state. 
The supercooled state can persist to a minimum temperature $T_1$, which may be zero.

Another important quantity of a first-order phase transition is the difference in the trace of the energy-momentum tensor 
between phases, which is the energy available for conversion to shear stress and so dictates the power of the gravitational wave signal. This is quantified in a dimensionless transition strength parameter $\alpha$, which we define below. 
We first note that the plasma enthalpy $w$, pressure $p$, and energy density $e$ are all related by $w=e+p$. We also introduce a useful quantity $\theta$ that is proportional the trace of the energy-momentum tensor:
\begin{equation}
\theta_{s,b} = \frac14 \left(e_{s,b}-3p_{s,b}\right) \ ,
\end{equation}
where the s/b subscripts represent quantities in the symmetric and broken phase, respectively.\footnote{We use the terms ``symmetric'' and ``broken'' for the two phases, following the convention in gauge theories. As we are considering cooling through the transition, the symmetric phase is the `old' phase and the broken phase is the `new' phase.}
The transition strength parameter is then defined as
\begin{equation}\label{alpha}
\alpha = \frac43 \frac{(\theta_s-\theta_b)}{w_s} \ .
\end{equation}
Another quantity which is closely related is the latent heat, found at $T_c$ by
\begin{equation}
L = e_{s}(T_c)-e_{b}(T_c) = 4(\theta_{s}(T_c)-\theta_{b}(T_c)),
\end{equation}
with the second equality following from the definition of the critical temperature, $p_b(T_c) = p_s(T_c)$. If the latent heat is comparable to the radiation energy density of the universe, we call the transition strongly first order. In terms of the transition strength, this happens when $\alpha\sim1$. We also call $\alpha \sim 0.1$ intermediate, and $\alpha \gg 1$ very strong, following \cite{Hindmarsh:2015qta,Hindmarsh:2017gnf,Cutting_2020}. The parameter $\alpha$ is a primary focus in this paper, as it can be directly accessed through a holographic calculation, and we will expand on it later.

Bubble walls are assumed to expand at a constant speed $v_w$ \cite{Steinhardt:1981ct}, which is determined by how the wall interacts with the surrounding plasma in the interplay between bubble expansion and frictional forces \cite{Liu:1992tn,Moore:1995si}. Fluid friction is thought to prevent runaway acceleration in phase transitions in gauge theories, although the details of the interactions between the particles of the plasma and the wall are under debate 
\cite{Bodeker:2017cim,Hoeche:2020rsg,Vanvlasselaer:2020niz}.
The wall speed is of particular importance as it impacts the kinetic energy production, and hence the 
gravitational wave power. 

Another important parameter is the transition rate  
\begin{equation}
\left.\beta \equiv \frac{\mathrm{d}}{\mathrm{d} t} \log \left(\frac{\Gamma(t)}{\mathcal{V}}\right)\right|_{t=t_{f}} \ ,
\end{equation}
where $\Gamma(t)/\mathcal{V}$ is the nucleation rate per unit volume in the symmetric phase. This is evaluated at a time $t_f$ which is at the temperature where the nucleation rate averaged over the whole universe peaks, and can be used to define the nucleation temperature \cite{Hindmarsh:2020hop}.  From these quantities the scale of the theory in the form of the typical bubble separation is set by
\begin{equation}\label{R_star}
R_* \propto\frac{v_w}{\beta} \ .
\end{equation}
The proportionality factor is an O$(1)$ number, which specifically for weak transitions is $(8\pi)^{1/3}$. As it is not known what the factor is for all transition strengths, we will use this number as a first approximation.

Finding $\beta$ involves a calculation of the effective action for 
non-constant fields, which is a straightforward 
procedure in a weakly coupled theory, but in a holographic set-up is challenging enough to merit a separate treatment. 
Holographic methods for calculating $v_w$ in this theory do not yet exist. 
When studying gravitational wave production we will therefore treat them 
as free parameters.\footnote{In weakly coupled theories, there are interesting correlations between $\beta$ and $\alpha$ \cite{Ellis:2018mja,Eichhorn:2020upj}.}  
For studying gravitational wave power spectra the more useful scale-setting combination is $R_*$.

\section{Holographic setup}\label{sec:holo}

The gauge/gravity duality is a powerful tool to deal with strongly coupled gauge systems and their phase structure, as strongly coupled systems on one side can be translated into weakly coupled systems on the other. The duality provides a ``holographic dictionary" which describes an exact linkage between quantities on the $d$-dimensional field theory side to quantities on the $(d+1)$-dimensional gravitational side, with surprising success in areas such as heavy ion collisions \cite{CasalderreySolana:2011us,Ramallo:2013bua,Brambilla:2014jmp}. 
Using the duality we are able to calculate quantities relevant for gravitational wave production in
phase transitions, which would otherwise be very hard to compute.

The model consists of gravity coupled to a bulk scalar field, with the following action
\begin{equation}
S_{\text{non-reg}} = \frac{2}{\kappa_5^2}\int \text{d}^{5}x\sqrt{-g}\left(\frac{\mathcal{R}}{4}-\frac{1}{2}\partial_{\mu}\phi\partial^{\mu}\phi-V(\phi)\right)+\frac{1}{\kappa_5^2}\int_{\partial\mathcal{M}}\text{d}^{4}x\sqrt{-\gamma}K \ ,
\end{equation}
where the first term is the (4+1)-dimensional Einstein-Hilbert action and the last term is the Gibbons-Hawking-York boundary term~\cite{PhysRevD.15.2752,PhysRevLett.28.1082}, with $\gamma$ representing the determinant of the induced metric on the boundary and $K$ giving the trace of the extrinsic curvature. The potential for the scalar field $V(\phi)$ is defined in terms of a superpotential $W(\phi)$ which was introduced in this holographic setting by \cite{super}. It is worth pointing out that the system is not expected to be supersymmetric and invoking the superpotential is merely a mathematical trick which allows to find solutions by solving a simpler set of first order equations   \cite{Freedman:2003ax}. By analogy with supersymmetric systems we we will dub the solutions to the first order system as ``BPS''  (Bogomol'nyi-Prasad-Sommerfield).

The general formula for the potential is as follows
\begin{equation}\label{pot}
V(\phi) = -\frac{4}{3}W(\phi)^2+\frac{1}{2}W'(\phi)^2 \ .
\end{equation}
The superpotential is chosen as in \cite{Bea:2018whf}, so as to provide the system with a first-order phase transition. It is dependent upon two parameters that we will specify in the numerical calculation (namely $\phi_M$ and $\phi_Q$), and has the form
\begin{equation}
LW(\phi) = -\frac32-\frac{\phi^2}{2}-\frac{\phi^4}{4\phi_M^2}+\frac{\phi^6}{\phi_Q} \ .
\end{equation}
When $W$ is inserted into the equation for the potential \eqref{pot}, one obtains
\begin{equation}\label{eq:scalarpotential}
\begin{aligned}
L^2 V(\phi) &= -3-\frac{3\phi^2}{2}-\frac{\phi^4}{3}-\left(\frac{1}{3\phi_M^2}-\frac{1}{2\phi_M^4}+\frac{2}{\phi_Q}\right)\phi^6 \\ &-\left(\frac{1}{12\phi_M^4}+\frac{6}{\phi_M^2\phi_Q}-\frac{4}{3\phi_Q}\right)\phi^8
+\left(\frac{2}{3\phi_M^4\phi_Q}+\frac{18}{\phi_Q^2}\right)\phi^{10}-\frac{4\phi^{12}}{3\phi_Q^2}.
\end{aligned}
\end{equation}
It is evident that both the potential and superpotential have a maximum at $\phi=0$. At the maximum the second derivative of the potential (which determines mass) takes the value $m^2=-3/L^2$. Following the usual holographic dictionary \cite{Aharony_2000} for a massive scalar in $AdS_5$, the field $\phi$ is dual to a scalar operator $\mathcal{O}$ with a scaling dimension determined by
\begin{equation}
\Delta(\Delta-4) = m^2L^2 \ .
\end{equation}
The larger solution determines the scaling dimension of the dual operator $\Delta_+=3$. The smaller solution corresponds to a coupling with dimension $\Delta_-=1$. We have deferred details of the holographic renormalisation to Appendix~\ref{app:holoreno} and hence speed forward to discussing the solutions of the model instead.

\begin{figure}[h]
  \centering
      \includegraphics[width=1\textwidth]{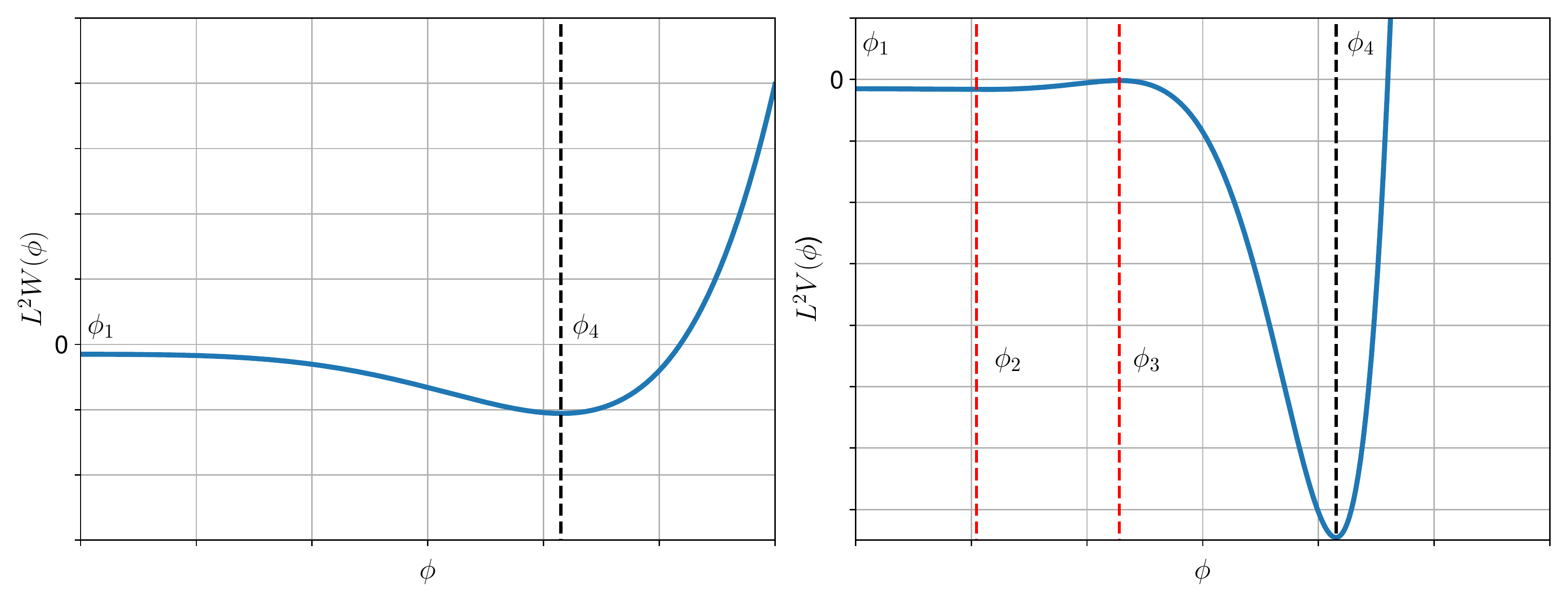}
  \caption{Graphs of the superpotential $W(\phi)$ and potential $V(\phi)$ of the theory, with interesting points shown. The superpotential and potential share the BPS extrema $\phi_1$ and $\phi_4$, whilst only the potential has the non-BPS extrema $\phi_2$ and $\phi_3$. In this plot the values used are $\phi_M=0.5$ and $\phi_Q=5.0$.}
  \label{fig:potentials}
\end{figure}

The potential and superpotential share a minimum and a maximum, with the minimum known as the BPS vacuum, but for some values of the parameters the potential also contains two ``non-BPS'' extrema in between these points not present in the superpotential, as seen in Fig.~\ref{fig:potentials}. The non-BPS minimum corresponds to a vacuum which persists to zero temperature. Some values that realise this situation are the ones used in \cite{Bea:2018whf}, $\phi_M\simeq0.5797$ and $\phi_Q=10.0$; see also related study in \cite{Gursoy:2018umf}. At zero temperature there can be solutions interpolating between the $\phi=0$ maximum close to the boundary and either the BPS or the non-BPS minimum in the deep interior of the geometry. Each of them correspond to different vacua of the dual field theory. At finite temperature there is a competition between the phases associated to each of these vacua, but even for values of the parameters where the non-BPS extrema are absent from the potential and there is a unique vacuum at zero temperature, a new phase appears at large enough temperatures. We will be interested in this last situation.

\subsection{Black brane solutions}

In order to determine the thermodynamic properties of this system, we need to find a family of black brane solutions. They can be  parametrised by the horizon value of $\phi$ (denoted $\phi_h$), with $\phi_h$ approaching the value where the potential has a minimum at lower temperatures and zero at higher temperatures.
The zero temperature solutions for $\phi$ indicate that we can use the scalar field as a coordinate, and so we employ the same metric choice as \cite{Attems_2016}, which can be expressed in the Eddington-Finkelstein form as 
\begin{equation}\label{Met:1}
ds^2 = e^{2A(\phi)}\left(-h(\phi)d\tau^2+d\bold{x}^2\right)-2e^{A(\phi)+B(\phi)}Ld\tau d\phi \ .
\end{equation}
The Einstein equations for this metric Ansatz are
\begin{equation}\label{EEs}
\begin{aligned}
A''(\phi)-A'(\phi)B'(\phi)+\frac23 &= 0 \\
h''(\phi)+(4A'(\phi)-B'(\phi))h'(\phi) &= 0 \\
\frac32 A'(\phi)h'(\phi)+(6A'(\phi)^2-1)h(\phi)+2e^{2B(\phi)}L^2V(\phi) &= 0 \\
4A'(\phi)-B'(\phi)+\frac{1}{h(\phi)}\left(h'(\phi)-e^{2B(\phi)}L^2V'(\phi)\right) &=0 \ .
\end{aligned}
\end{equation}
We observe that our scalar field is bounded by $0\leq\phi\leq\phi_h$, with the requirement that the blackening factor goes to zero at the horizon $h(\phi_h)=0$. 

We emulate the master function procedure first introduced in \cite{Master}, where by combining the Einstein equations and derivatives thereof we can reduce the problem to only depend on the ``master function'' and the potential. We consider a smooth ``generating function'' which will be related to our metric components by
\begin{equation}\label{GtoA}
G(\phi) = \frac{dA(\phi)}{d\phi} \ .
\end{equation}
Replacing this in the field equations and manipulating them we find
\begin{equation}
\frac{G^{\prime}(\phi)}{G(\phi)+\frac{4 V(\phi)}{3 V^{\prime}(\phi)}}=\frac{d}{d \phi} \log \left(\frac{G^{\prime}(\phi)}{G(\phi)}+\frac{1}{6 G(\phi)}-4 G(\phi)-\frac{G^{\prime}(\phi)}{\left(G(\phi)+\frac{4 V(\phi)}{3 V^{\prime}(\phi)}\right)}\right) \ ,
\end{equation}
leaving us with a second order non-linear differential equation to solve. To reduce this to first order equations more suitable for numerical integration, we introduce a new variable defined as
\begin{equation}
H(\phi) = \frac{G'(\phi)}{G(\phi)} \ ,
\end{equation}
which when entered into the master equation, and after some further manipulation, yields two differential equations to be solved:
\begin{equation}
G'(\phi) = G(\phi)H(\phi)
\end{equation}
and
\begin{equation}
H'(\phi) = \frac{H(\phi)}{\left(1+\frac{4\gamma_{1}(\phi)}{3G(\phi)}\right)}\left[2H(\phi)+\frac{2}{G(\phi)}+\frac{8\gamma_{1}(\phi)}{9G^2(\phi)}+\frac{1}{\gamma_{2}(\phi)}+4G(\phi)\left(1+\frac{4\gamma_{1}(\phi)}{3G(\phi)}\right)\right] \ .
\end{equation}
Here we have set 
\begin{equation}
\gamma_{1}(\phi) = \frac{V(\phi)}{V'(\phi)}, \qquad \gamma_{2}(\phi) = \frac{V'(\phi)}{V''(\phi)}, \qquad \gamma_{3} = \frac{V''(\phi)}{V'''(\phi)},
\end{equation}
for brevity, with the last definition preemptively added. Further following the procedure of \cite{Master}, the next step is to find the series solution of the master equation around the horizon $\phi_h$, which translates to finding series solutions for both $G(\phi)$ and $H(\phi)$. By requiring that the blackening factor goes to zero at the horizon, {\emph{i.e.}}, $h(\phi_h) = 0$, we can find an expression for $G(\phi_h)$ by combining the last two of the Einstein equations in (\ref{EEs}) and evaluating them at the horizon. Derivatives of the expression before horizon evaluation can give an expansion up to any desired order. Taylor expanding around $\phi_h$ therefore gives (denoting $\gamma(\phi_h)=\gamma^h$)
\begin{equation} \label{hor:1}
G(\phi) = -\frac{4}{3}\gamma_1^h\left[1+\frac{1}{2}(\phi-\phi_h)\left(\frac{\gamma_2^h-\gamma_1^h}{\gamma_1^h\gamma_2^h}\right)\right]+\textrm{O}(\phi-\phi_h)^2
\end{equation}
and
\begin{equation} \label{hor:2}
H(\phi) = \frac{\gamma_2^h-\gamma_1^h}{2\gamma_1^h\gamma_2^h}\left[1+\frac{2}{3}(\phi-\phi_h)\left(1+\frac{\gamma_{1}^h}{\gamma_2^h\gamma_3^h}\frac{(\gamma_3^h-\gamma_2^h)}{(\gamma_2^h-\gamma_1^h)}-\frac{8}{3}\gamma_1^h\right)\right]+\textrm{O}(\phi-\phi_h)^2 \ ,
\end{equation}
with the condition for $H$ at the horizon  
\begin{equation}
\frac{dH}{d\phi}\Bigg|_{\phi_h} = H(\phi_h)\left(\frac{2}{3}  \frac{\gamma_{1}^h}{\gamma_2^h\gamma_3^h}\frac{(\gamma_3^h-\gamma_2^h)}{(\gamma_2^h-\gamma_1^h)}-\frac{16 \gamma _1^h}{9}-\frac{3}{2 \gamma _1^h}-\frac{4}{3}\right)\ .
\end{equation}
We also wish to know what is happening for these quantities at the other boundary in our model, where $\phi \rightarrow0$. Expansion for small $\phi$ of (\ref{hor:1}) and (\ref{hor:2}) gives a simple leading behaviour 
\begin{equation}
G(\phi) = \frac{dA(\phi)}{d\phi} = -\frac{1}{\phi}+\ldots\ ,
\end{equation}
and
\begin{equation}
H(\phi) = -\frac{1}{\phi} + \ldots \ .
\end{equation}
Once the master function is determined, the other metric quantities have a simple dependence on it.

The first relation comes immediately from the definition of $G(\phi)$ (\ref{GtoA}), that we integrate to obtain $A(\phi)$ 
\begin{equation}\label{met1}
A(\phi) = -\log\left(\frac{\phi}{\Lambda L}\right)+\int^{\phi}_{0}\left(G(\varphi)+\frac{1}{\varphi}\right)d\varphi \ .
\end{equation}
Rearranging the first of our field equations (\ref{EEs}) for $B'(\phi)$ and integrating gives us
\begin{equation}\label{met2}
B(\phi) = \log(|G(\phi)|)+\int^{\phi}_{0}\frac{2d\varphi}{3G(\varphi)} \ .
\end{equation}
Finally, eliminating $h'(\phi)/h(\phi)$ from the last two field equations in (\ref{EEs}) leaves $h(\phi)$ in terms of known quantities, taking the form
\begin{equation}\label{met3}
h(\phi) = -\frac{e^{2B(\phi)}L^2(4V(\phi))+3G(\phi)V'(\phi)}{3G'(\phi)} \ .
\end{equation}
With our differential equations and metric functions specified and our boundary conditions established in the form of horizon quantities (\ref{hor:1}) and (\ref{hor:2}), we now show how this master function can be solved.

Analytic solutions to our system of equations are rare, only occurring for specially selected master functions/potentials (see Ref.~\cite{Chamblin} where $G(\phi) = -1/(3\gamma)$). Therefore, as we are searching for specific solutions of a relatively complicated potential, we will need to resort to numerical methods (see Appendix~\ref{app:numerics}).

\section{Thermodynamics}\label{sec:thermo}

The entropy and temperature in the dual field theory are determined by the Bekenstein-Hawking entropy and Hawking temperature of the black brane. The entropy is proportional to the area of the horizon while the temperature is proportional to the surface gravity. These can be expressed in terms of metric components as
\begin{equation}
s = \lim_{\phi\to\phi_h}\frac{2\pi}{\kappa_5^2}\sqrt{(g_{xx})^3}, \qquad T = \lim_{\phi\to\phi_h}\frac{1}{2\pi}\frac{\partial_{\phi}\sqrt{g_{\tau\tau}}}{\sqrt{g_{\phi\phi}}} \ .
\end{equation}
Using the metric (\ref{Met:1}), we can read off the entropy density and temperature as follows
\begin{equation}
s = \frac{2\pi}{\kappa_5^2}e^{3A(\phi_h)}, \qquad LT = \frac{e^{A(\phi_h)-B(\phi_h)}}{4\pi}|h'(\phi_h)| \ .
\end{equation}

All of these functions are now readily evaluated at the horizon using the formulae (\ref{met1})-(\ref{met3}) found in Section~\ref{sec:holo}. Evaluating $h'(\phi)$ at the horizon simply requires the use of either of the last two field equations in (\ref{EEs}). Combining everything together, we find the entropy density and temperature in terms of the master function
\begin{equation}
s = \frac{2\pi}{\kappa_5^2}\left(\frac{\Lambda L}{\phi_h}\right)^3\text{exp}\left\{3\int^{\phi_h}_{0}\left(G(\phi)+\frac{1}{\phi}\right)d\phi\right\}
\end{equation}
and
\begin{equation}
T = -\Lambda\frac{L^2V(\phi_h)}{3\pi\phi_h}\text{exp}\left\{\int^{\phi_h}_{0}\left(G(\phi)+\frac{1}{\phi}+\frac{2}{3G(\phi)}\right)\right\} \ .
\end{equation}
At zero temperature the theory becomes conformal at the fixed points (UV and IR). We then expect the entropy density to tend to $\frac{2\pi^2}{45}g_{*}T^3$ close to those fixed points, where $g_*$ is the effective number of relativistic degrees of freedom of the corresponding CFT. Defining a dimensionless and rescaled measure of 
the entropy
\begin{equation}
\left(\frac{s}{T^3}\right)_R = \frac{\kappa_5^2}{2\pi^4 L^3}\frac{s}{T^3} = -\left(\frac{3}{L^2V(\phi_h)}\right)^3\text{exp}\left(-\int^{\phi_h}_{0}\frac{2d\phi}{G(\phi)}\right) \ ,
\end{equation}
we now have an expression purely depending on the master function and the potential. We compare our calculation of $s/T^3$ with that obtained in \cite{Bea:2018whf} for particular values of the parameters in the potential ($\phi_M\simeq0.5797,\phi_Q=10.0$) in Appendix~\ref{app:numerics}. Conformal symmetry is achieved at high temperature, when the coupling to the scalar operator is negligible compared with the temperature, and we approach the solution $\phi\sim 0$ in the gravity dual. Here our rescaled quantity tends to 1, and so this implies that our number of degrees of freedom on the gravity side is 
\begin{equation}
\frac{s}{T^3} = \frac{2\pi^4 L^3}{\kappa_5^2} \ ,
\end{equation}
which depends on the radius of curvature $L$ and the five dimensional Newton's constant $\kappa_5^2=8\pi G_5$.

\begin{figure}
  \centering
      \includegraphics[width=1\textwidth]{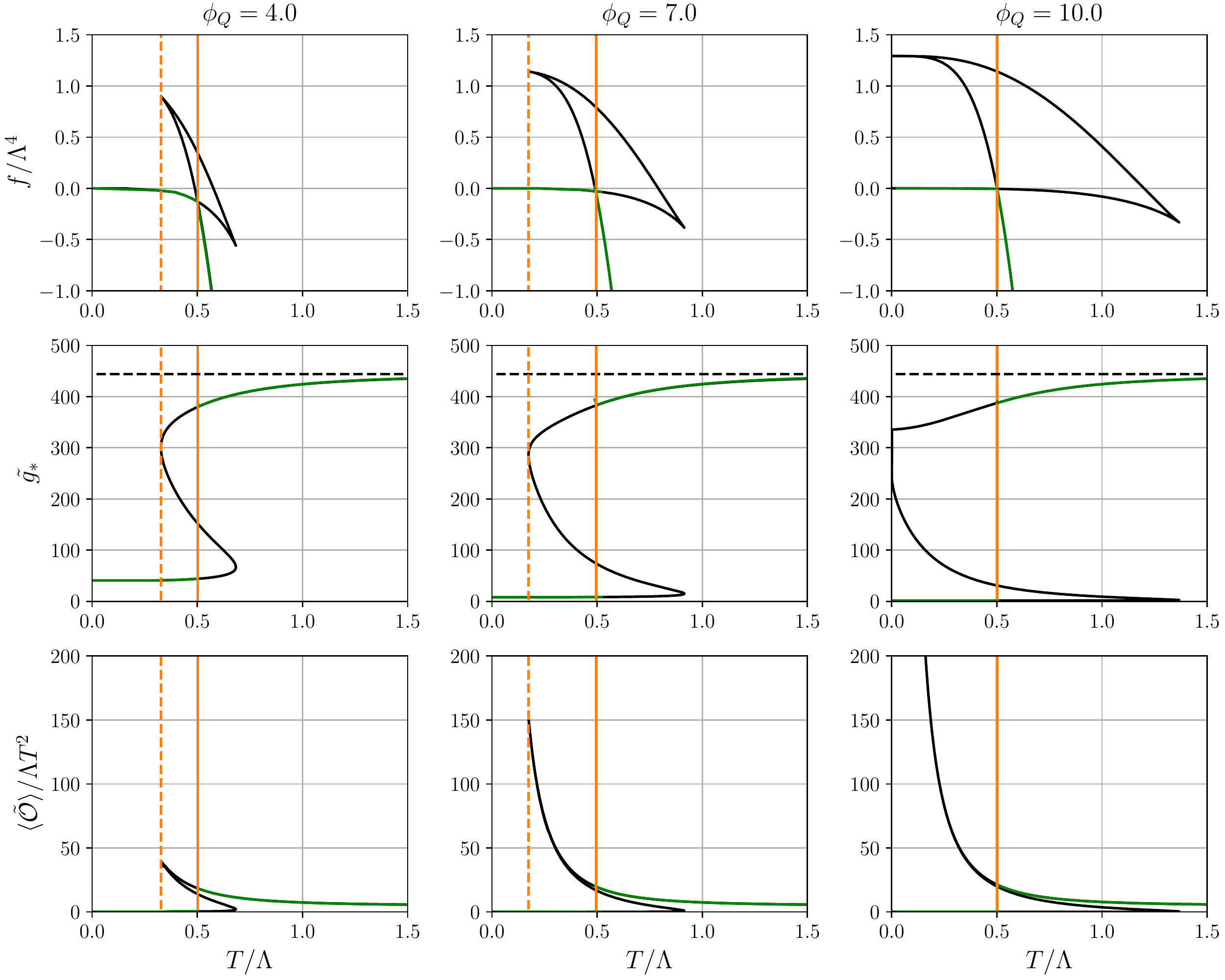}
  \caption{Free energy density $f$, rescaled degrees of freedom $\tilde{g}_*$, and dimensionless scalar condensate $\langle\tilde{\mathcal{O}}\rangle$ over temperature for varying $\phi_Q$ at constant $\phi_M\approx0.58.$ The solid yellow line on each plot shows the critical temperature $T_c$, the dashed yellow line on each plot shows the last temperature at which the metastable phase exists $T_1$, the green line shows the stable phase, and the black dashed line on the middle row plots shows the asymptotic value of $\tilde{g}_*$.}
  \label{fig:grid}
\end{figure}

\begin{figure}
  \centering
      \includegraphics[width=1\textwidth]{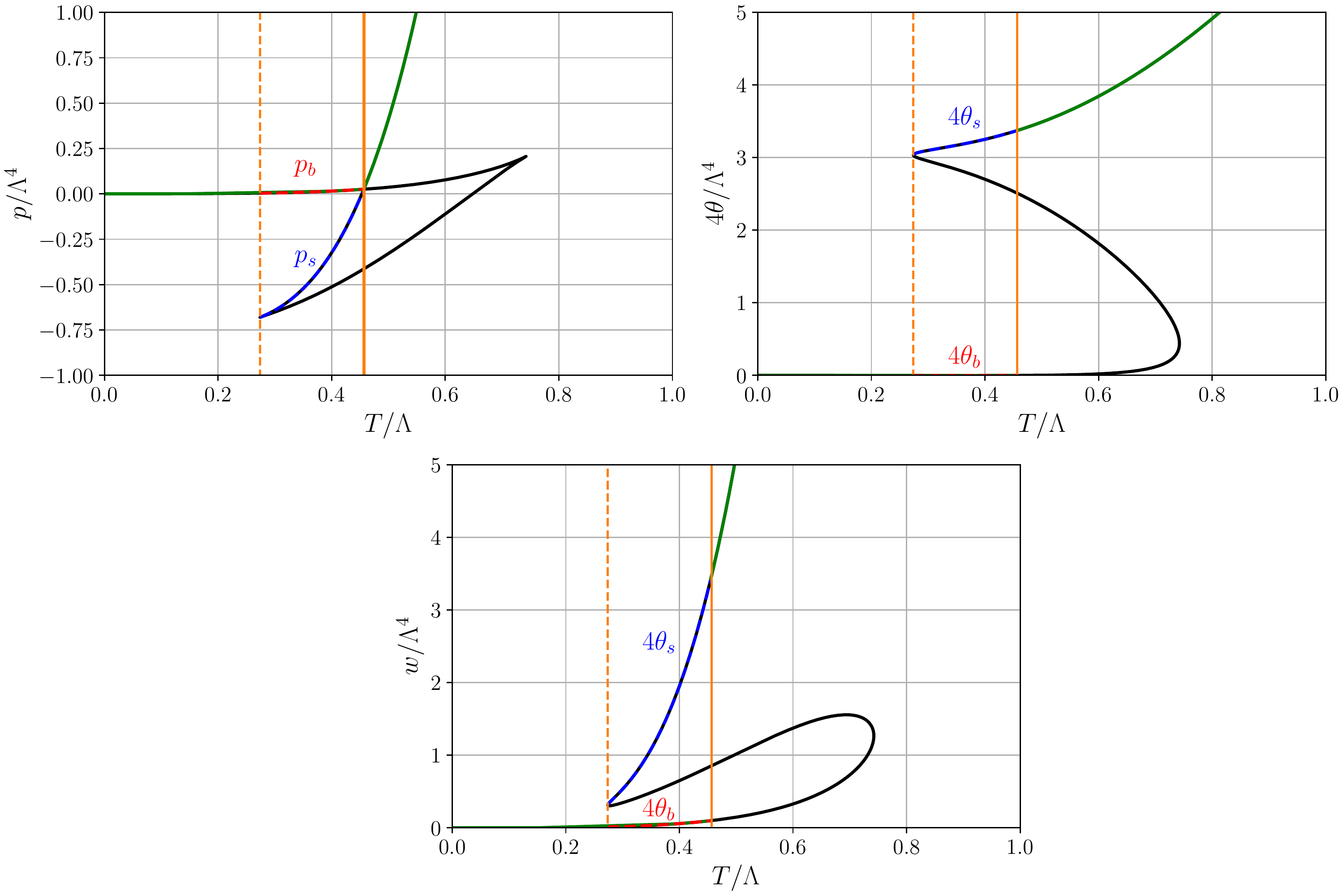}
  \caption{Pressure, trace of the energy-momentum tensor, and enthalpy density for $\phi_Q=10.0$ and $\phi_M=0.7$, with the different branches labeled. The green line shows the stable phase, the solid yellow line shows $T_c$, and the dashed yellow line shows $T_1$.}
  \label{fig:branches}
\end{figure}

For the study of the phase transition, we will also need the energy density and pressure, 
$e$ and $p$. The pressure can be obtained directly from $s$ and $T$ using the thermodynamic derivative
\begin{equation}
s = \frac{dp}{dT},
\end{equation}
which is easily integrable (numerically) to give 
\begin{equation}
p = \int^{T}_{0}s(\tilde{T})d\tilde{T} \ .
\end{equation}
This is consistent with the holographic renormalisation analysis of Appendix~\ref{app:holoreno}, in that the vacuum contribution vanishes.
The energy density is obtained through the thermodynamic relation
\begin{equation}
e = Ts-p \ .
\end{equation}
Integrating numerically in this way introduces errors, 
which we checked by comparing to the correct $T^4$ behaviour at high temperature and found matches well. 

Another quantity we are interested in is the expectation value of the scalar operator. As the energy-momentum tensor can be written as $T_{\mu\nu} = \text{diag}(e,p,p,p)$, we may take the trace and use the Ward identity\footnote{For this particular model a possible contribution to the trace anomaly $\sim\Lambda^4$ vanishes, see, {\emph{e.g.}}, \cite{Hoyos:2016cob,Ecker:2017fyh} and Appendix \ref{app:holoreno}.}
\begin{equation}
\langle T_{\mu}^{\mu}\rangle = -\Lambda\langle\mathcal{O}\rangle
\end{equation}
to write
\begin{equation}
-\langle T_{\mu}^{\mu}\rangle = e-3p = \Lambda\langle\mathcal{O}\rangle \ .
\end{equation}
We will fix units to $\kappa_5^2/L^3=1$, so implicitly we are computing rescaled quantities such as
\begin{equation}
\langle\tilde{\mathcal{O}}\rangle = \frac{\kappa_5^2}{L^3}\langle{\mathcal{O}}\rangle \ ,
\end{equation}
and similarly for the thermodynamic potentials. We plot the expectation value of the scalar operator, as well as the free energy ($f=-p$) and the rescaled effective degrees of freedom $\tilde{g}_*=g_*{\kappa_5^2}/{L^3}$ for various values of $\phi_Q$ at fixed $\phi_M$ in Fig.~\ref{fig:grid}. 
We first see that, as $\phi_Q$ increases, $f$ moves away from its usual ``swallow tail" first-order transition shape and the energy density and free energy tend to the case with non-BPS extrema, \emph{i.e.} with both phases persisting down to zero temperature. In the IR, the scalar field is non-zero and will be most relevant to all physics considerations.
In the UV region, however, all operators tend to $\propto\Lambda T^2$ and to very similar numeric values as well (no difference up to the 13th decimal place for these examples). This is explained by noticing that in the UV region we are considering the vicinity of $\phi\to0$, which results in the potential acting as
\begin{equation}
V(\phi)\approx -3+\mathrm{O}(\phi^2) \ ,
\end{equation}
independent of $\phi_M$ or $\phi_Q$ values.

Recalling our definition for $\alpha$ (\ref{alpha}) we now see that we have everything necessary for its calculation, except for knowing how to split the energy density and the free energy into their broken and unbroken phase sections. To do so we remember that the two different branches of the free energy that cross each other on the free energy plot correlate to the two different phases in question, and so the quantities we need are the sections of these branches which exist simultaneously before the critical temperature $T_c$ as shown in Fig.~\ref{fig:branches}. The critical temperature is defined as the temperature at which this crossing happens, in which it becomes energetically favourable to transition from one phase to the other.

\section{Parameter scanning}\label{sec:parameterscanning}

With all definitions and calculation techniques set in place we can finally move to scanning over the holographic parameters to see how varying these will change the quantities relevant for gravitational wave spectra. The two ``dials" we can turn in this theory are the parameters in the potential, $\phi_M$ and $\phi_Q$; varying these changes the shape of the potential and therefore the black brane solutions and thermodynamic quantities derived from them. Increasing $\phi_M$ effectively means bringing the two non-BPS extrema in the potential $V(\phi)$ closer together, until at a certain value for each $\phi_Q$ these merge as an inflection point and then disappear completely. 
The approximate equation of the region with non-BPS extrema is 
\begin{equation}
\label{e:nonBPSline}
\phi_Q \gtrsim 150\phi_M^5, 
\end{equation}
which was found by a fit to the numerical solution of the equations $V'(\phi) = 0$ and $V''(\phi) = 0$. 

We have chosen to explore parameter ranges without these non-BPS extrema, 
as they produce a theory with a metastable minimum at $T=0$.  
These are unattractive for cosmological model-building, as the Universe could instead be trapped in an eternally inflating phase. 

\begin{figure}[h]
  \centering
      \includegraphics[width=1\textwidth]{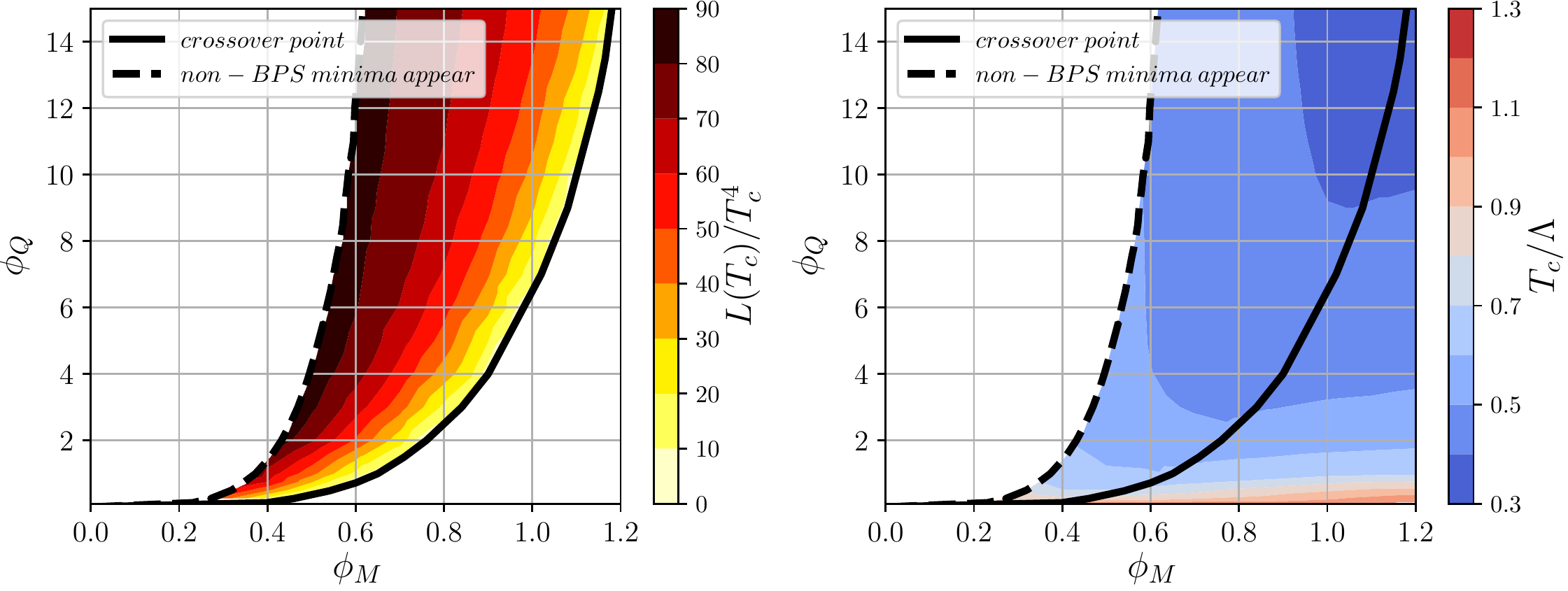}
  \caption{Filled contours of latent heat $L$ in units of the critical temperature $T_c$ for different values of the two scalar potential parameters on the left, and of the critical temperature $T_c$ in units of the coupling for different values of the scalar potential parameters on the right. The contours found past the crossover line on the critical temperature plot are from the peak of the interaction measure.}
  \label{fig:critical}
\end{figure}

Fig.~\ref{fig:critical} shows the latent heat and the critical temperature of the phase transition over a 
region in the $(\phi_M, \phi_Q)$ plane.  
The boundary of the region with non-BPS extrema is marked with a dashed line.
Where there is a first-order transition, $T_c$ it is defined as the temperature at which the free energy in both phases is equal. In the cross-over region $T_c$ is defined as the temperature in which the ratio of the trace of the stress energy tensor to the enthalpy (also known as the interaction measure $I$) peaks, where
\begin{equation}
I = \frac{e-3p}{e+p}.
\end{equation}

Increasing $\phi_M$ away from the region (\ref{e:nonBPSline}), the latent heat of the first-order transition decreases until it vanishes, as seen in Fig.~\ref{fig:critical}, at which point the theory presumably undergoes a second-order phase transition. 
The region of cross-overs has the approximate formula 
\begin{equation}
\label{e:1OPTlim}
\phi_Q \lesssim  6.5 \phi_M^5 ,
\end{equation}
obtained by a numerical fit. The boundary is marked with a solid line in Fig.~\ref{fig:critical}. 
Increasing $\phi_Q$ however has the opposite effect, but much more slowly. As $\phi_Q$ grows the system is pushed into a stronger first-order phase transition with higher latent heat.


\begin{figure}[h]
  \centering
      \includegraphics[width=0.7\textwidth]{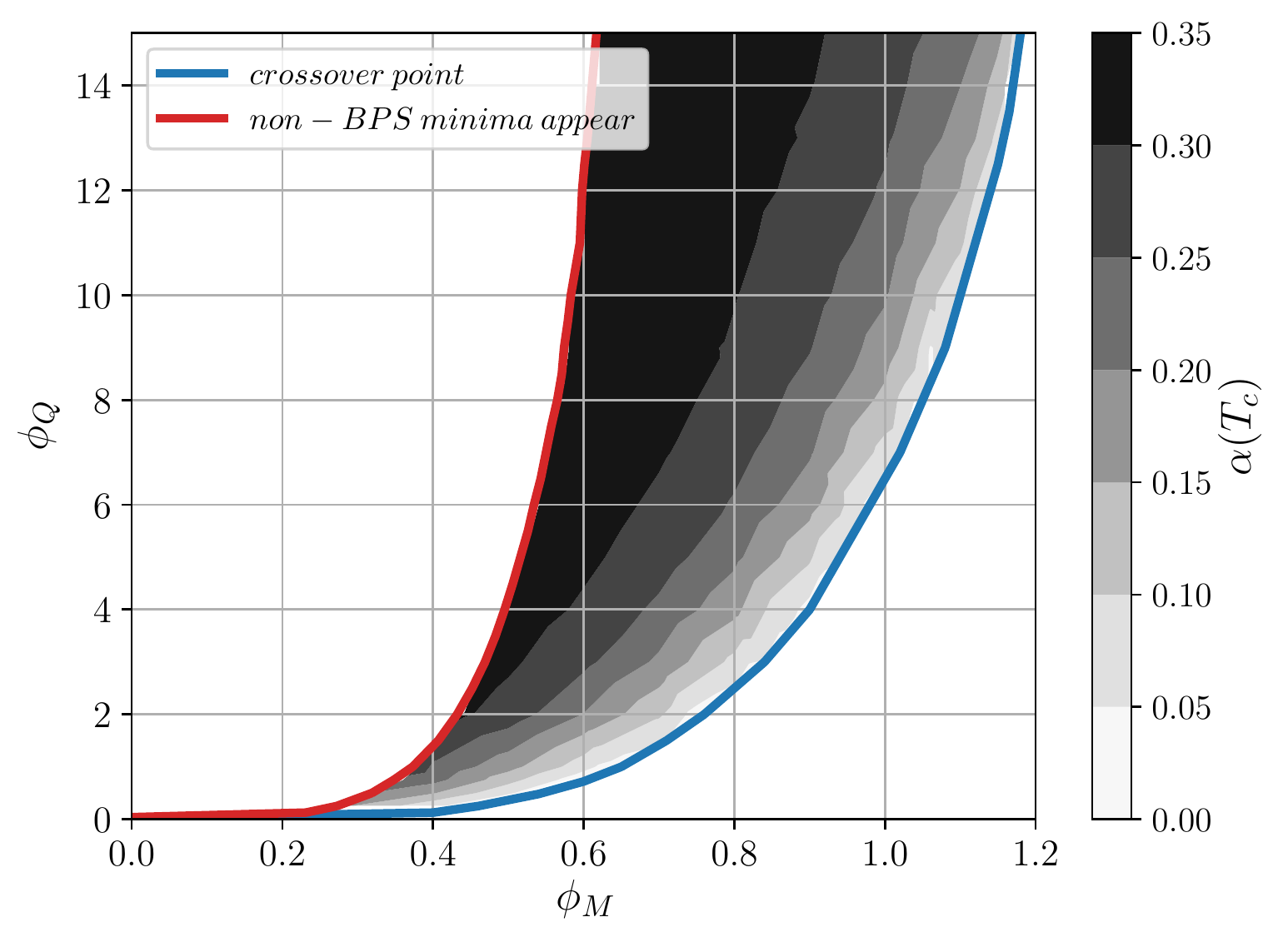}
  \caption{Scan of the transition strength parameter $\alpha$ at the critical temperature for different values of the scalar potential parameters.}
  \label{fig:alpha}
\end{figure}

The measure of the strength of the phase transition relevant for gravitational production is $\alpha$,  defined around Eq.~\ref{alpha}. 
In Fig.~\ref{fig:alpha} we show the value of $\alpha(T_c)$, the value at the critical temperature, 
in the $(\phi_M,\phi_Q)$ plane.
It is very promising that there are systems accessible with intermediate transition strengths, as it is thought that, for $\alpha\sim0.1$ or greater, a signal will be observable at space-based detectors, as we will discuss in the following section. 

\begin{figure}[b]
\centering
    \includegraphics[width=1\textwidth]{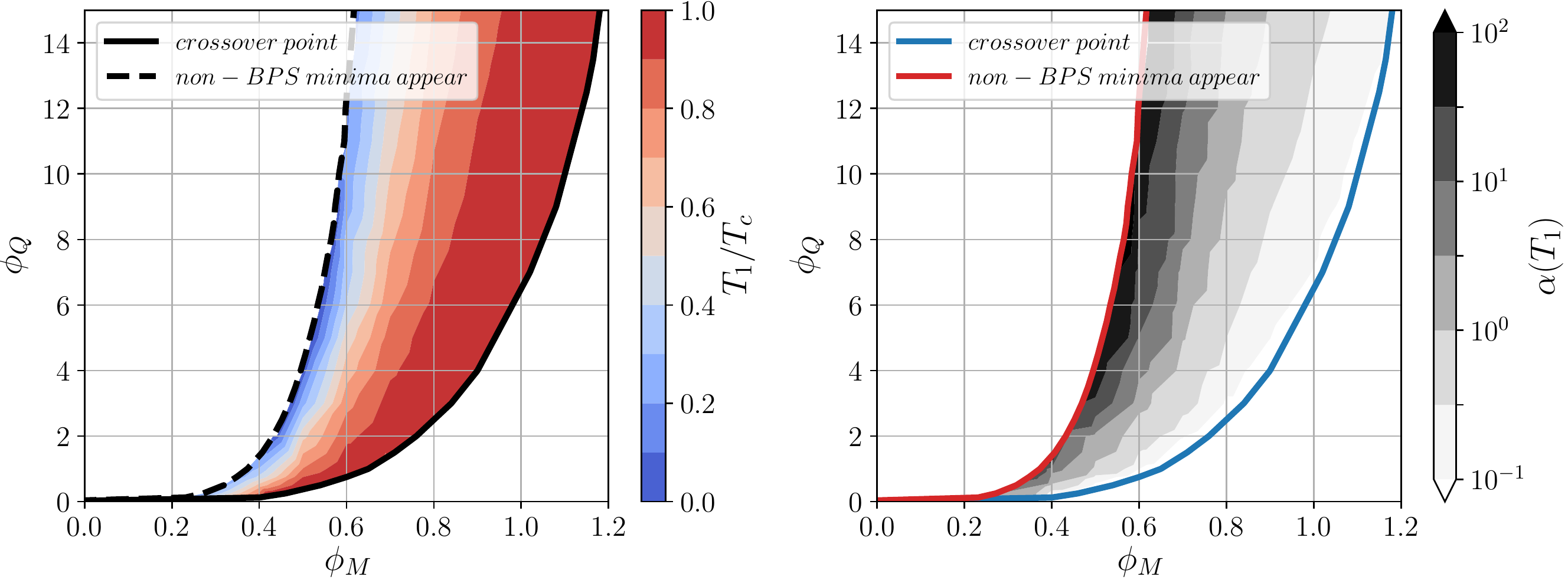}
\caption{Left: contours of $T_1/T_c$ in the $(\phi_M,\phi_Q)$ plane, 
where $T_1$ is the lowest temperature at which the 
metastable phase exists. Right: transition strength parameter $\alpha$ at 
temperature $T_1$.}
\label{fig:pair}
\end{figure}

\begin{figure}[h]
  \centering
      \includegraphics[width=0.7\textwidth]{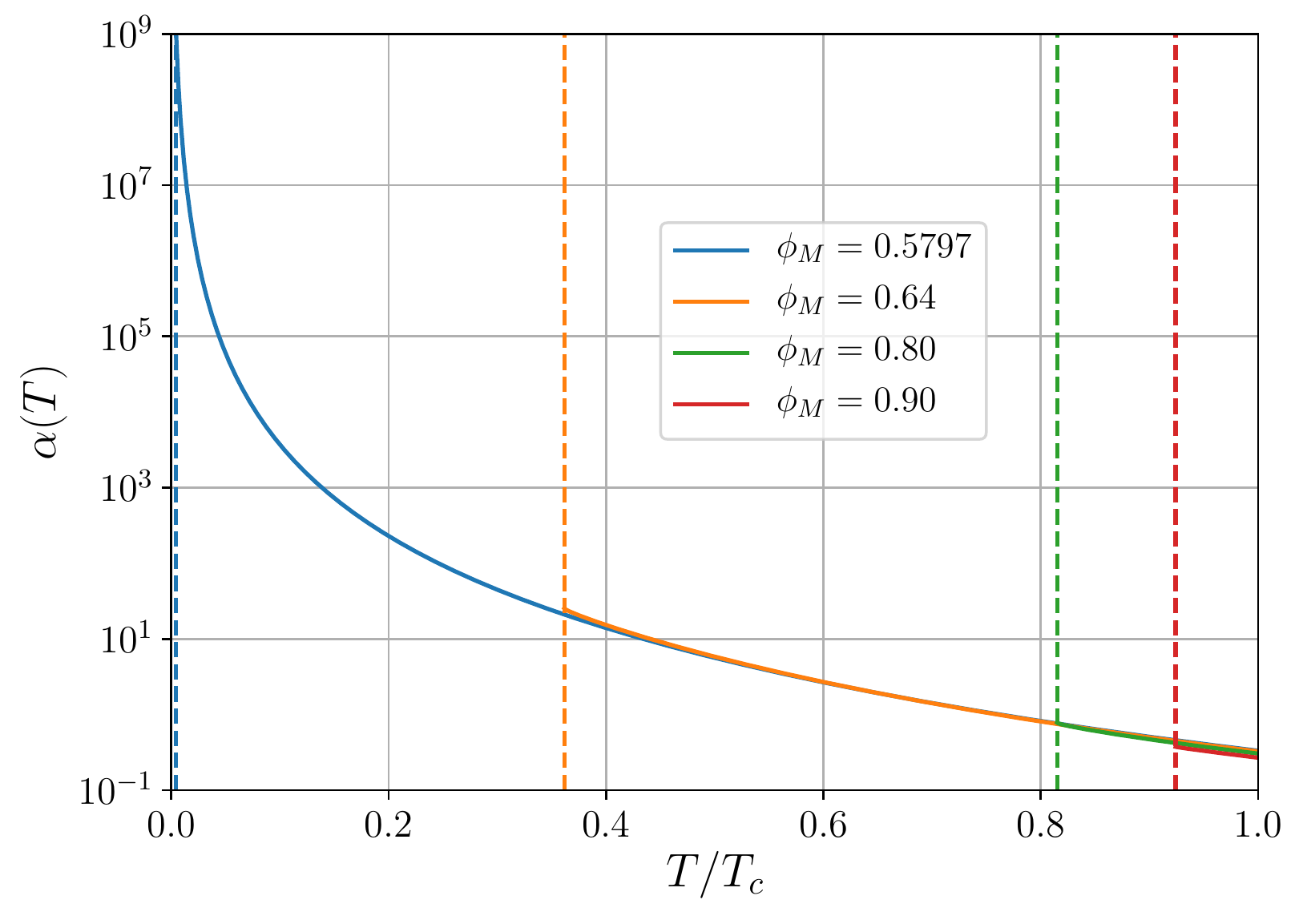}
  \caption{Temperature dependence of the transition strength parameter $\alpha$ in terms of $T/T_c$ for $\phi_Q=10.0$. The dashed lines represent at which value the original, high-temperature minimum is no longer separated by a potential barrier and so disappears, as seen in Fig.~\ref{fig:quartic}.}
  \label{fig:logalphas}
\end{figure}

The most relevant quantity for the strength of the phase transition is 
$\alpha$ at the nucleation temperature $\alpha(T_N)$. 
We have not established the nucleation temperature of the transition, 
but we do know that the lowest nucleation temperature is the lowest temperature at which the 
metastable phase exists, which we denote $T_1$. 
In Fig.~\ref{fig:pair} we show the ratio $T_1/T_c$, showing the maximum possible supercooling,
as well as $\alpha(T_1)$. 
As $\alpha$ increases below the critical temperature, as 
shown in Fig.~\ref{fig:logalphas}, 
$\alpha(T_1)$ represents the maximum value of the transition strength parameter achievable by supercooling. 
At the boundary where the non-BPS minima appear $T_1 \to 0$, and the enthalpy of the metastable phase can reach arbitrarily low values. We therefore see diverging values of $\alpha(T_1)$ 
near that boundary, which can also be seen in the curve for $\phi_M\simeq0.5797$ in Fig.~\ref{fig:logalphas}.

\begin{figure}[h]
  \centering
      \includegraphics[width=1\textwidth]{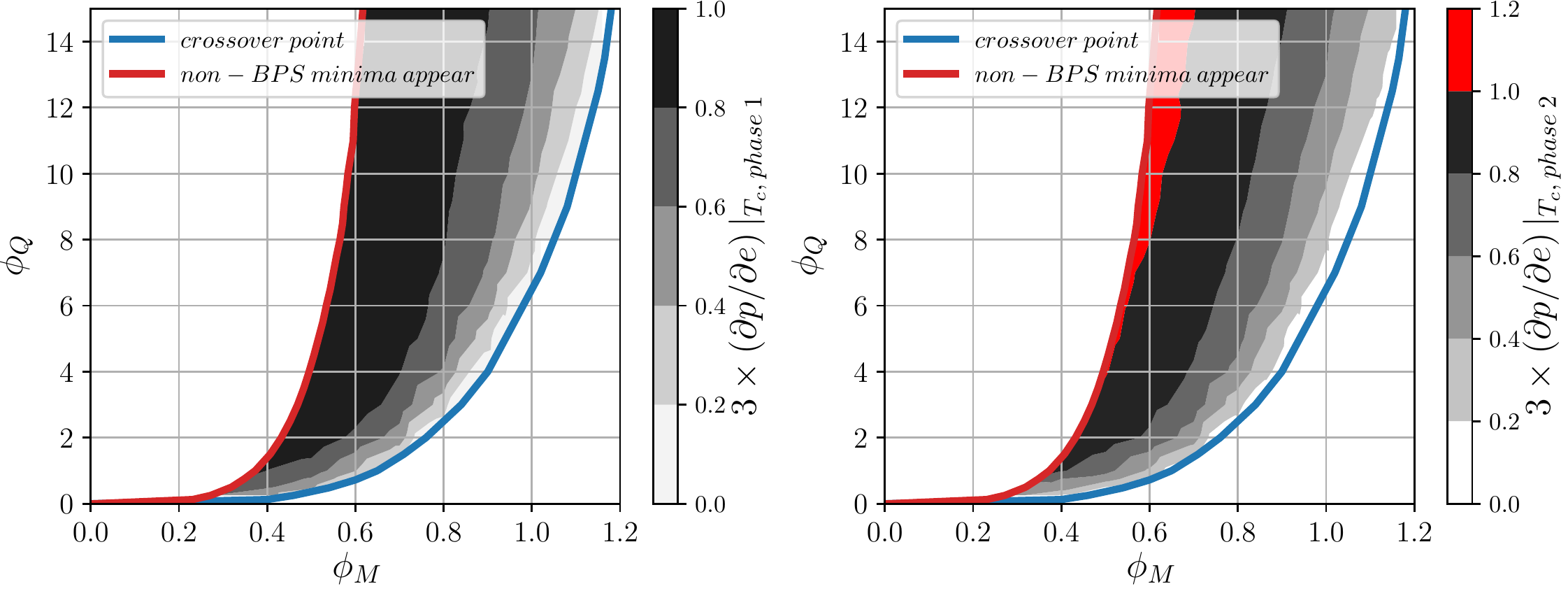}
  \caption{Scan of $3\times\pa p/\pa e$ at the critical temperature in both phases, with phase one
  (the symmetric phase) on the left and phase two (the broken phase) on the right. 
  Values that fall above the conformal stiffness value $\pa p/\pa e = 1/3$ are shown as a red filled contour.}
  \label{fig:stiffscans}
\end{figure}

The final parameter scans in Fig.~\ref{fig:stiffscans} are that of the stiffness of the equation of state $\pa p/\pa e$ in both phases, a quantity which for barotropic fluids can be identified as the square of the sound speed $c_s^2$. 
We have not properly established with a fluctuation analysis that the speed of sound is indeed the square root 
of the stiffness, but we will nevertheless denote $\pa p/\pa e = c_s^2$ in the following and use both interchangeably.
As can be seen in the figure, the conformal value of $3\times\pa p/\pa e = 1$ is reached in the region where the transition strength is strongest and in fact goes above the conformal value in phase two for the strongest transitions, however, the stiffness steadily declines to relatively small values as we journey towards a crossover transition. The effect this could have on the signal is has been recently discussed in \cite{Giese_2020,giese2020modelindependent} and is 
described in Appendix~\ref{app:omegagw}.

\section{Gravitational waves}\label{sec:GW}

To see what the results from the holographic model imply for the detectability of gravitational waves by the LISA mission, we turn to the gravitational wave spectra calculations. 
The quantity of interest is $h^2\Omega_{\mathrm{gw}}(f)$, the energy density contained in gravitational waves relative to the total per log frequency interval, sometimes called just the power spectrum.
This is to be compared to the detector noise, which is quoted in terms of a sensitivity 
$h^{2} \Omega_{\mathrm{Sens}}(f)$, which is the gravitational wave power spectrum producing the 
same amplitude signal as the detector noise at frequency $f$.  The comparison is through the 
signal-to-noise ratio (SNR) defined below.

The base model for these spectra is summarised in 
\cite{Caprini:2019egz}, based on numerical simulations \cite{Hindmarsh:2017gnf} and a physical understanding 
in terms of sound waves \cite{Hindmarsh:2013xza,Hindmarsh_2019}. 
There have been developments in understanding since the appearance of \cite{Caprini:2019egz},
concerning the finite lifetime of the source \cite{guo2020phase}, 
and kinetic energy production in strong phase transitions \cite{Cutting_2020}, 
and those with stiffness away from 1/3 \cite{Giese_2020}.

%

With these improvements (see Appendix \ref{app:omegagw}), the power spectrum can be written 
\begin{equation}
h^2\Omega_{\mathrm{gw}}=2.061 h^2 F_{\mathrm{gw}, 0} \left(H_{\mathrm{N}} R_{*}\right)K^{2} \tilde{\Omega}_{\mathrm{gw}} C\left(\frac{f}{f_{\mathrm{p}, 0}}\right)
\Bigg(1-\frac{1}{\sqrt{1+2x}}\Bigg) \Sigma \ ,
\end{equation}
where $h$ is the Hubble parameter today with value $0.678 \pm 0.009$ \cite{2016Planck}, $H_N$ is the Hubble rate at nucleation, $F_{\mathrm{gw}, 0}$ is an attenuation factor 
\begin{equation}
F_{\mathrm{gw}, 0} = (3.57\pm0.05) \times 10^{-5}\left(\frac{100}{g_*}\right)^{\frac13} \ ,
\end{equation}
$K$ is the kinetic energy fraction of the fluid around the expanding bubbles of the stable phase, 
and $x=H_{\mathrm{N}} R_{*} / \sqrt{K}$, where $R_*$ is the mean bubble separation. 
The final factor $\Sigma$ takes into account the effect of 
kinetic energy suppression recently observed in strong phase transitions \cite{Cutting_2020}.

We use $R_* = (8\pi)^{1/3}v_w/\beta$ though the relation is accurate only for weak transitions ($\alpha \lesssim {\rm O}(10^{-2})$). We do not yet have a good theory of the function $R_*(v_w,\beta)$; 
replacing $(8\pi)^{1/3}$ by the sound speed as advocated in \cite{Caprini_2020} is likely to lead to overestimating the power.

The constant $\tilde{\Omega}_{\mathrm{gw}}$ has a numerically-determined value of order $10^{-2}$ \cite{Hindmarsh:2017gnf}.  We take it to be $1\times10^{-2}$, which replicates the correct peak amplitude for an intermediate strength transition with $v_w = 0.92$ but under-predicts the power spectrum for other transitions, so is a conservative estimate. Lastly, $C(s)$ is the spectral shape function 
\begin{equation}
C(s) = s^3\left(\frac{7}{4+3s^2}\right)^{7/2} \ 
\end{equation}
with peak frequency
\begin{equation}\label{eq:peakfrequency}
f_{p,0} \simeq 26.2\left(\frac{1}{H_N R_*}\right)\left(\frac{T_N}{100\text{GeV}}\right)\left(\frac{g_*}{100}\right)^{1/6}\mu \text{Hz} \ 
\end{equation}
which comes from fits around the peak of the numerically-determined GW power spectrum. 

As can be seen, the peak frequency of the power spectrum depends on the nucleation temperature $T_N$, and also on the wall velocity $v_w$ and transition rate $\beta$ through $R_*$ (equation \ref{R_star}). 
The peak power is controlled by $R_*$ and the kinetic energy fraction, which in turn is 
sensitive to the transition strength parameter $\alpha$ and the wall speed $v_w$, and the stiffness.
Hence the power spectrum is controlled by 
all four crucial parameters mentioned in Section~\ref{sec:holo} as well as the sound speed $c_s$ \cite{Giese_2020,giese2020modelindependent}. 

We take $\alpha$ to be equal to its holographic value and therefore neglect the Standard Model degrees of freedom, which is discussed in more detail at the end of this section. To properly calculate the nucleation temperature would require a full derivation of the nucleation rate through the effective action, which is beyond the scope of this paper. To circumvent this we realise that the nucleation temperature will always be lower than the critical temperature, and by Fig.~\ref{fig:pair} we see that lowering the temperature will in fact increase the transition strength, thereby increasing the gravitational wave signal. We therefore take a conservative estimate of the nucleation temperature as being at the critical temperature, $T_N = T_c$. 

This now allows us to translate the temperatures we found in previous sections which are in units of the coupling to physical temperatures in units of GeV. 
The lack of new physics up to the TeV scale motivates placing a lower bound of 1 TeV on the coupling $\Lambda$. 
From Fig.~\ref{fig:critical} we can estimate that this will limit our nucleation temperature to a range from $300$ GeV to $1.3$ TeV. The vast majority of parameter values (approximately three quarters) fall within the range of temperatures $400 - 600$ GeV, which in turn motivates the choice of a nucleation temperature of $T_N = 500$ GeV when plotting spectra from here on. 

\begin{figure}[h]
  \centering
     \includegraphics[width=0.85\textwidth]{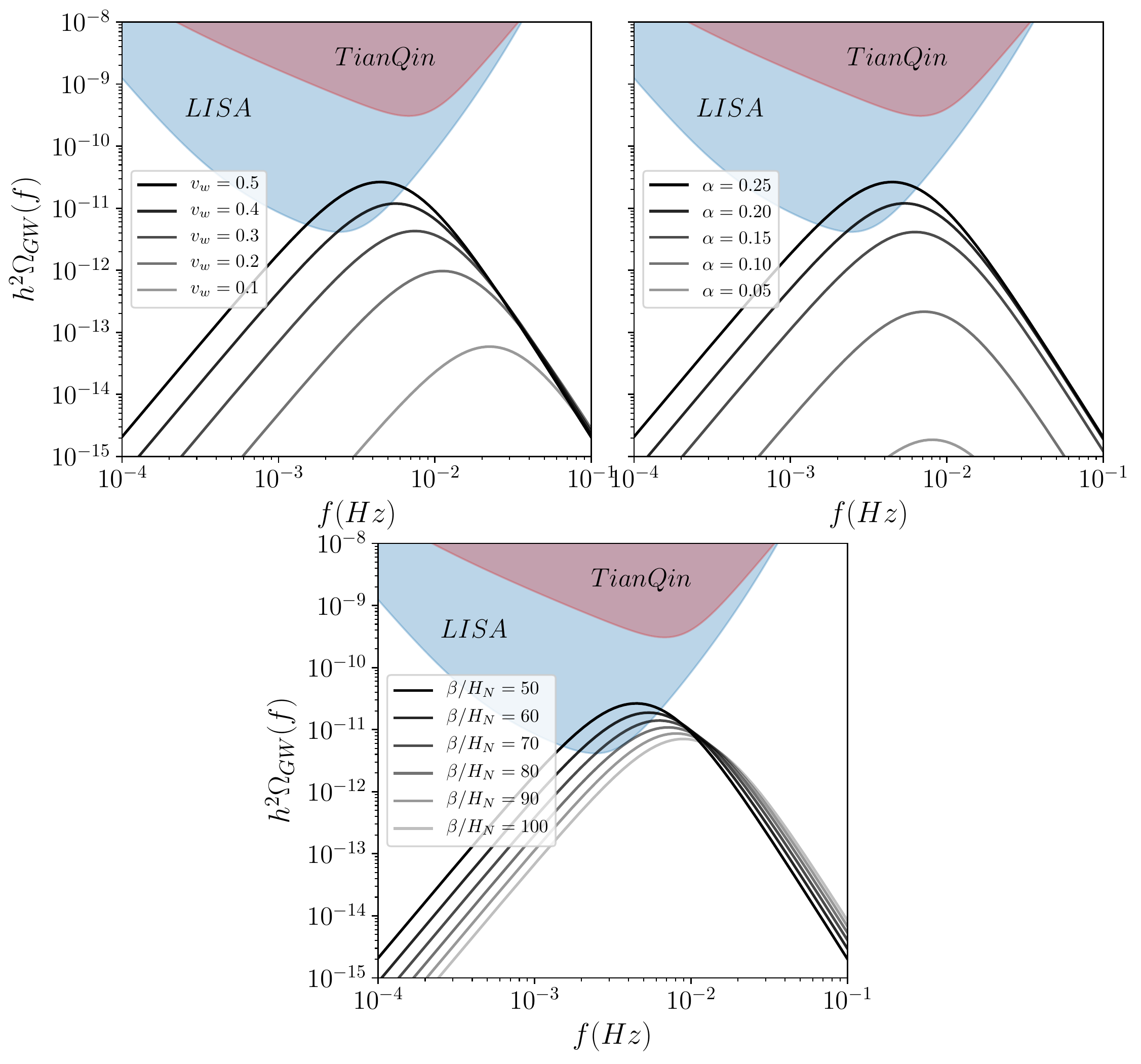}
  \caption{Power spectra for varying $v_w$, $\alpha$, and $\beta/H_N$ respectively. The baseline takes values $g_* = 100$, $\alpha = 0.25$, $\pa p/\pa e = \alpha$, $\beta/H_N = 50$, $v_w = 0.5$, and $T_N = 500$GeV, and is depicted as the darkest line on every plot.}
  \label{fig:spec3}
\end{figure}

The only parameter in the gravitational wave power spectrum which depends on this choice is the peak frequency $f_{p,0}$. By inspecting (\ref{eq:peakfrequency}) we note that an increase of $T_N$ will result in a shift to higher frequencies of the gravitational wave power spectral curves, without a change in shape. 
For parameter values of $g_*=100, \alpha=0.25, \beta/H_N=50, v_w=0.5$ used in Fig.~\ref{fig:spec3} values of $T_N$ up to around $\sim 750$ GeV will keep the peak within the LISA sensitivity window. 


A gravitational wave detector's sensitivity to cosmological sources is determined from the instrumental noise, converted into an equivalent gravitational wave signal 
\begin{equation}
h^{2} \Omega_{\mathrm{Sens}}(f)=\frac{2 \pi^{2}}{3 H_{0}^{2}} f^{3} S_{n}(f) \ ,
\end{equation}
where $S_n(f)$ is the noise power spectral density and $H_0$ is the Hubble rate today.

LISA's expected noise power spectral density is fully described in \cite{Scidoc}. 
Another mission Taiji is also planned with very similar parameters to LISA \cite{Ruan_2020}. 
A third planned space-based mission is TianQin, with a substantially different configuration \cite{Luo_2016, Wang_2020}.
In Fig.~\ref{fig:spec3} we display the model gravitational wave power spectra for varying $\beta/H_N$, $v_w$, and $T_N$ to demonstrate the strength of the signals from phase transitions relative to the sensitivity curves of LISA (taken to also apply to Taiji) and TianQin. 
We have chosen $g_*=100$ and $T_N = 500$ GeV for all graphs. The darkest line on every plot takes values of $\beta/H_N=50$, $v_w=0.5$, and $\alpha = 0.25$, with one variable shifting for each plot. 
We take the stiffness equal to $\alpha$, in view of the correlation shown in the next section.

Significant quantities determining whether the signal will be in the detectable range are the wall speed and transition strength parameter.
As it is evident from the figures, most of the signals we show are strong enough to be detected by LISA, if the peak frequency falls into its range of sensitivity, although all fall short of TianQin's direct detection level. Despite this, particular choices of higher wall speeds and stronger transitions than what we chose could quickly push the signal into the range that allows signal detection from both detectors. 

To more reliably indicate whether a signal will be seen by a mission we turn to the signal-to-noise ratio SNR. The SNR allows comparison of a gravitational wave signal with the detector's base noise level and whether the signals produced will be discernible from it. Once the power spectrum and detector sensitivity is known, the SNR follows from 
\begin{equation}
\mathrm{SNR}=\sqrt{\mathcal{T} \int_{f_{\min }}^{f_{\max }} \mathrm{d} f\left[\frac{h^{2} \Omega_{\mathrm{gw}}(f)}{h^{2} \Omega_{\mathrm{Sens}}(f)}\right]^{2}} \ ,
\end{equation}
where $\mathcal{T}$ is the mission duration time.


\begin{figure}[h]
  \centering
     \includegraphics[width=1\textwidth]{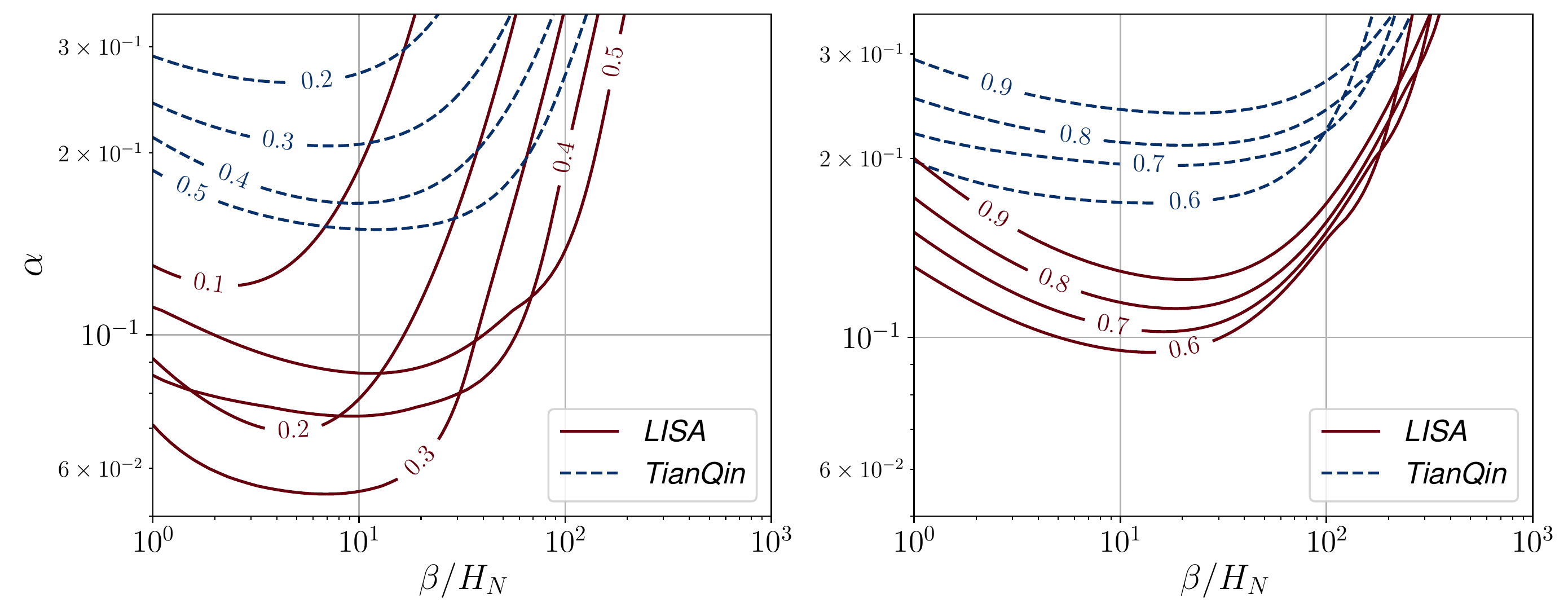}
  \caption{Curves of constant wall speed $v_w$ in the $\alpha-\beta$ plane for LISA and TianQin $\Omega_{\mathrm{gw,exp}}$, with dark red and dark blue lines representing the detectable signal-to-noise ratio limit of 10 for each wall velocity. The left plot displays wall speeds of $v_w = 0.5$ and below, and the right plot displays wall speeds of $v_w = 0.6$ and above, with the area contained above each line being theoretically detectable for that wall velocity, which is labeled on the line itself. The curve for wall velocity $v_w = 0.1$ for TianQin is not detectable in the range of transition strengths shown, and the temperature used for this plot is $T_N=500$GeV.}
  \label{fig:SNR}
\end{figure}

For the LISA mission the duration  is taken to be four years in orbit and a duty cycle of usable science data of 75\% making $t_{obs} = 3$yr. Displayed in Fig.~\ref{fig:SNR} is a plot combining both the SNR and three of the important parameters ($\alpha$, $\beta/H_N$, $v_w$) for both LISA and TianQin. We take that the stiffness $\pa p/\pa e$ for each point equal to the value of $\alpha$, a correlation which will be seen later in Fig.~\ref{fig:speedofsound}. We note, however, that we have not yet formally calculated the speed of sound in our model, though we expect it to lie close to the square root of the stiffness, at least for high temperatures. A proper calculation requires an analysis of the dispersion relations of the quasi-normal mode spectra of the coupled scalar field and metric components fluctuations, which we hope to perform elsewhere.

It is usually considered that a SNR above 10-20 is highly likely to be detectable \cite{Tinto_2017}, and so we produce contours of 10 and above for nine different wall speeds for each detector, which we split into two plots for clarity. 

\begin{figure}[h]
  \centering
     \includegraphics[width=1\textwidth]{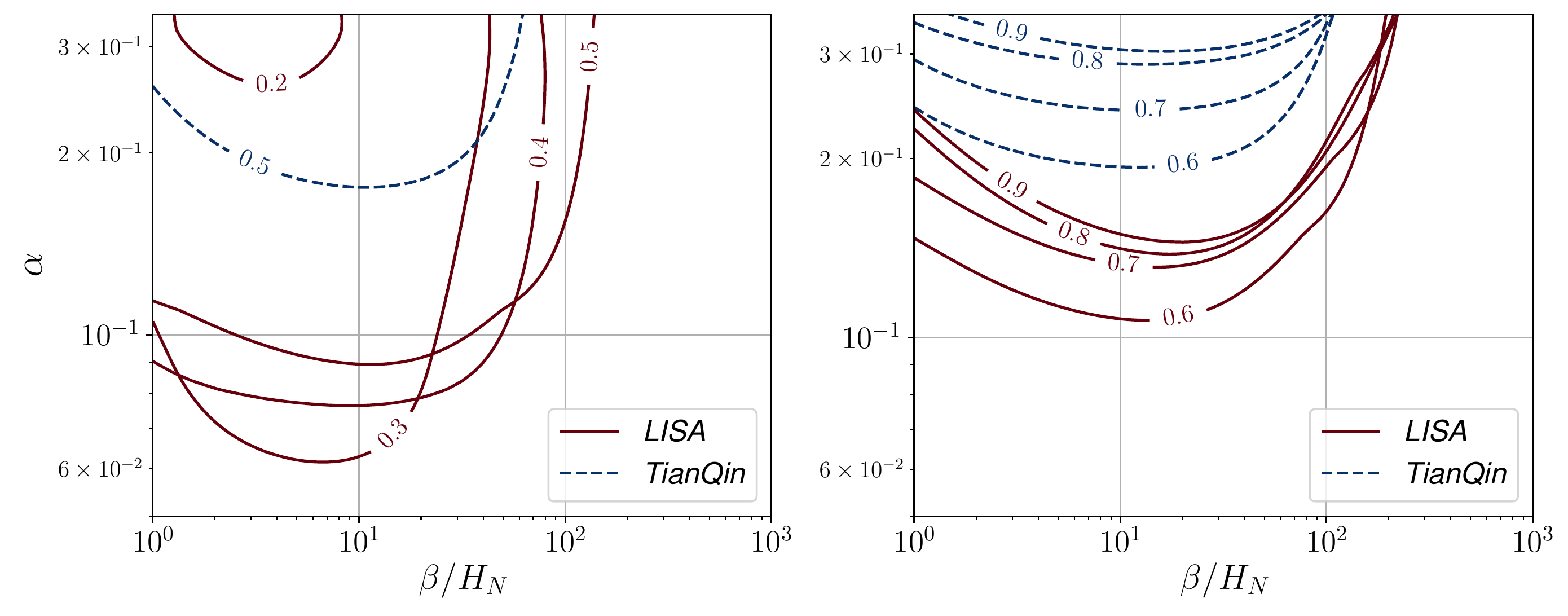}
  \caption{Curves of constant wall speed $v_w$ in the $\alpha-\beta$ plane for LISA and TianQin matching Fig.~\ref{fig:SNR}, now with the suppression $\Sigma$ applied giving $\Omega_{\mathrm{gw}}$, with dark red and dark blue lines representing the detectable signal-to-noise ratio limit of 10 for each wall velocity. Once again the left plot displays wall speeds of $v_w = 0.5$ and below, and the right plot displays wall speeds of $v_w = 0.6$ and above with the temperature used for this plot being $T_N=500$GeV.}
  \label{fig:SNR2}
\end{figure}

In Fig.~\ref{fig:SNR} we produce these signal-to-noise ratios incorporating the majority of the most recent discoveries including the finite lifetime of the source and the effect of varying sound speeds. In Fig.~\ref{fig:SNR2} we also incorporate the recently discovered suppression of kinetic energy production in strong phase transitions $\Sigma(v_w,\alpha)$, which as can be seen has a large effect on slower wall speeds, removing most contours for TianQin with slow wall speeds in our parameter range. We show with and without this modification as the data on this effect is sparse and therefore possibly not as illuminating, with some new points having not been analysed in great detail such as not being checked for lattice convergence.

Finally, we discuss the effect of the Standard Model degrees of freedom on the system. First, suppose the Standard Model and holographic degrees of freedom are at the same temperature. If so, there will be no change to the difference in vacuum energy, however there will be a contribution to the enthalpy, as $w_{tot} = w_{holo}+w_{SM}$. Including this into our definition for $\alpha$ modifies the value found as
\begin{equation}
\alpha_{tot} = \alpha_{holo}\left(1+\frac{\kappa_5^2}{L^3}\frac{g_*^{SM}}{g_{*,holo}^{sym}}\right)^{-1} \ ,
\end{equation}
which depends only on the ratio of degrees of freedom and the constants from the holographic model.

\begin{figure}[h]
  \centering
     \includegraphics[width=0.7\textwidth]{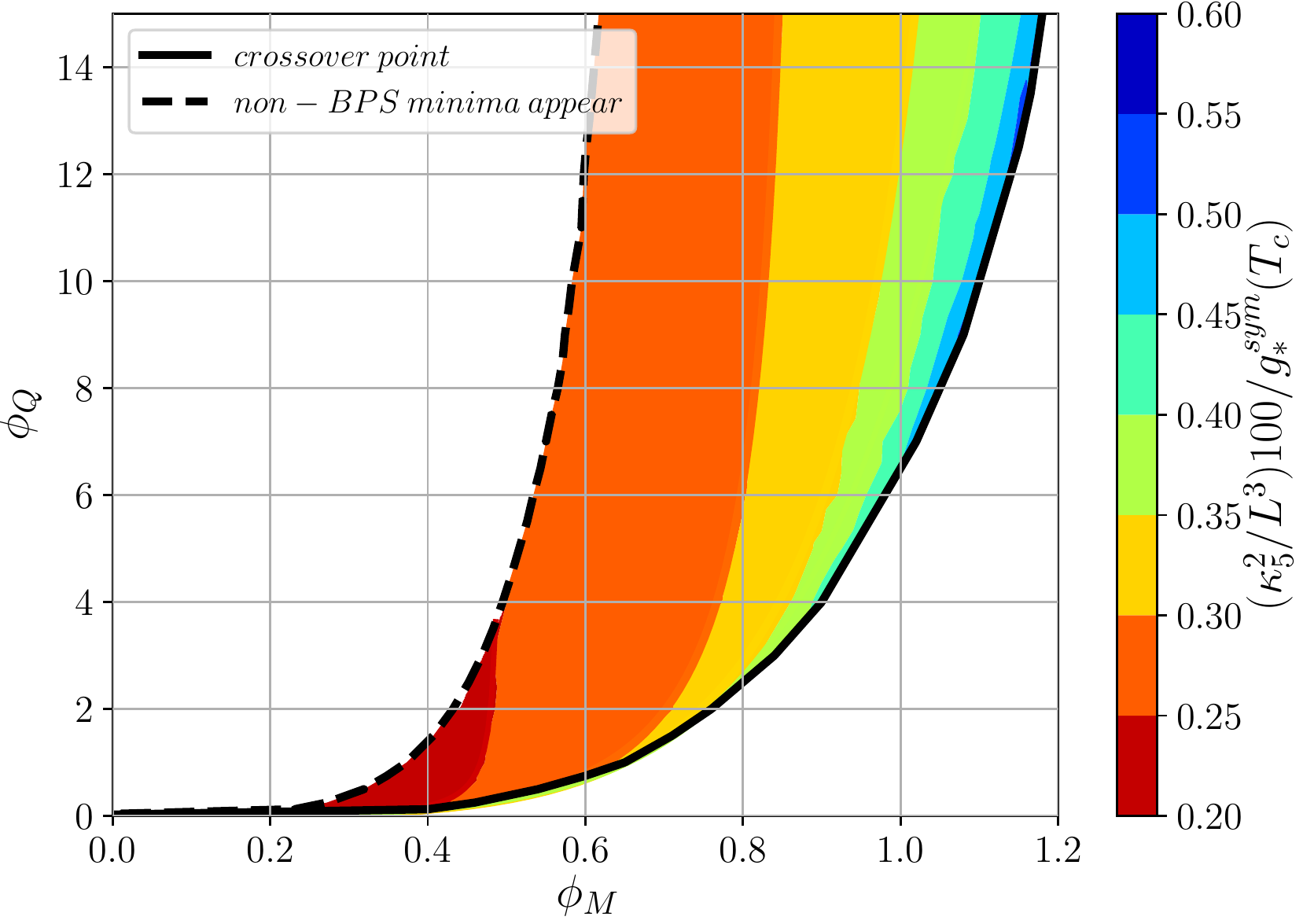}
  \caption{Contour plot of the ratio of a reference effective number of relativistic degrees of freedom  (100) with the holographic degrees of freedom, scaled by the holographic parameters $\kappa_5^2/L^3$, in the symmetric phase at the critical temperature.}
  \label{fig:dof}
\end{figure}

We see from Fig.~\ref{fig:dof} that the degrees of freedom ratio is small (always less than unity in units of $\kappa_5^2/L^3$), but it is greatly influenced by the gravity dual parameters. This in turn indicates that the ratio is contingent on the gauge group of the dual field theory $N^2$, which when large will render the standard model contributions negligible. The figure therefore informs how close to a large $N$ limit the theory is required to be in order to neglect standard model corrections. As holography is best formulated in a large $N$ limit however, neglecting this contribution is justified.

\section{Discussion}\label{sec:discussion}

In this paper we studied a particular model for phase transitions in a field theory, stemming from a putative strongly coupled sector described by holography. We adopted the so-called bottom-up gauge/gravity approach, using a 5-dimensional model with a single scalar field, whose bulk potential (\ref{eq:scalarpotential}) has two free parameters, the coefficients of the quartic and sextic terms of a superpotential \cite{Bea:2018whf}.  From a model builder's perspective this allows us the freedom to explore generic features without the greater complexity of a proper string theory construction.

We focused on the thermodynamic parameters relevant for gravitational wave production: the critical temperature, the latent heat, the transition strength parameter $\alpha$, the minimum temperature for metastability, and the stiffness $\pa p /\pa e$. We then explored the implications for gravitational wave production in the early universe, if the strongly coupled sector described by the model belonged to an extension of the Standard Model.

The theory has one dimensionful parameter $\Lambda$, which can be viewed as the coupling of a dimension-3 scalar operator $\mathcal{O}$ in the effective action of the field theory, and the scale of new physics.  We find that the theory has a first order phase transition in the approximate region given in Eq.~\eqref{e:1OPTlim}. 
Outside this region is a cross-over.

\begin{figure}[h]
  \centering
     \includegraphics[width=1\textwidth]{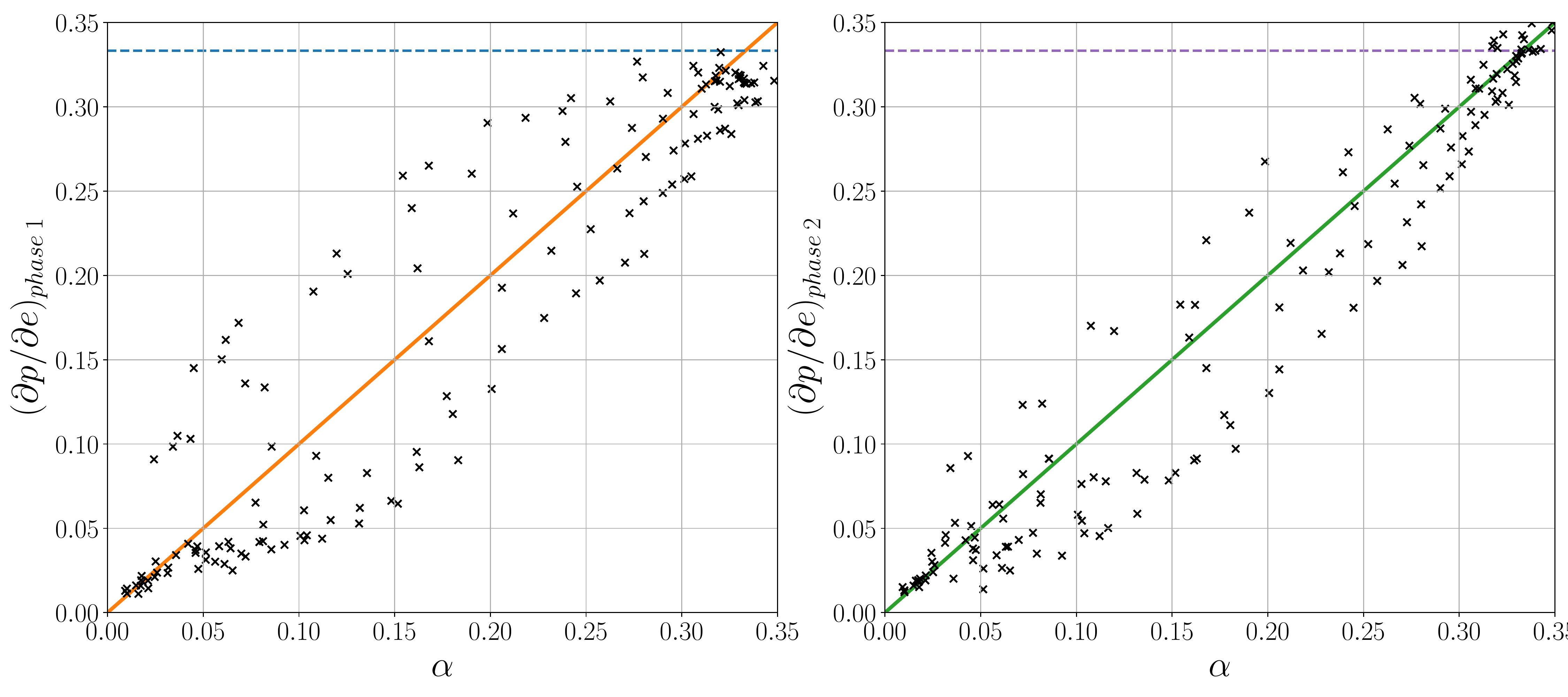}
  \caption{Scatter plots for the stiffness of the equation of state for phase one (the symmetric phase) on the left and phase two (the broken phase) on the right. The dashed line represents the conformal value $\partial p/\partial e = 1/3$. Density of points is not significant in this plot and is due to how the values of $\phi_M$ and $\phi_Q$ were sampled.}
  \label{fig:speedofsound}
\end{figure}

The critical temperature of the transition $T_\text{c}$ is O$(\Lambda)$, and generally $T_\text{c} < \Lambda$.  
We found the minimum temperature to which the metastable phase persisted $T_1$, finding that in the 
region given in Eq.~\eqref{e:nonBPSline} 
the metastable state persisted to zero temperature.  
The transition strength parameter at $T_\text{c}$ is generally in the range $0.1 \lesssim \alpha \lesssim 0.3$, although it 
drops to zero at the boundary with the cross-over region, as does the stiffness.  The stiffness can be larger than $1/3$, but is 
also generally in the range $0.1 \lesssim \pa p/\pa e \lesssim 0.3$.  It is strongly correlated with the transition strength 
(see Fig.~\ref{fig:speedofsound}).

Our study of gravitational waves was necessarily restricted by the lack of a prediction for the scale $\Lambda$, and the equilibrium nature of the calculation, which gave access only to the transition strength parameter and the stiffness. 
It is therefore illustrative in nature.  To gain insight into observational prospects, 
we assumed a critical temperature of $T_\text{c} = 500$ GeV, with negligible supercooling, 
and that the number of degrees of freedom of the Standard Model was small in comparison with that of the dual field theory.  

To calculate the gravitational wave power spectrum, we used the LISA Cosmology 
Working Group recipe \cite{Caprini_2020}, augmented by the kinetic energy suppression factor of Ref.~\cite{Cutting_2020} 
(see Appendix~\ref{app:omegagw}). 
We explored the consequences of stiffness away from the conformal value of 1/3 using the kinetic energy fraction algorithm 
recently presented in \cite{giese2020modelindependent}.

With this set of choices, we found that the transitions are strong enough to be observable at LISA over a wide range of parameter space, and there is also a smaller range detectable by TianQin.  
In order to be more definitive, it is important to calculate the speed of the phase boundary $v_\text{w}$, and the amount of supercooling, which is fixed by the rate of change of the tunnelling probability $\beta$. 
These are harder calculations, to which we will return in future.

It is interesting to speculate about which features we have found are generic in a strongly coupled transition.  That the critical temperature is below the masses of new states (unless there are also approximate symmetries broken at the transition) is a feature of QCD, in the generalised sense of the temperature of peak interaction measure, which is well below the nucleon and glueball masses. This is in contrast to weakly-coupled theories, where the masses of new states 
are generally down by a power of a coupling constant from transition temperature.  
One can also argue that the transition strength parameter is generally intermediate in strength ($\alpha \sim 0.1$). For example, in a confinement transition in a large $N$ theory, as O($N^2$) degrees of freedom become 
massive and are removed from the effective number of relativistic degrees of freedom, also O($N^2$).  
While the phase transition in the bottom-up theory we studied is not obviously a confinement transition in a real field theory, it shares this feature of a large change in the number of degrees of freedom, evident in Fig.~\ref{fig:grid}.

The strong departures of the stiffness from the conformal value of $1/3$ may also be generic, although its physical origin is difficult to understand. 
In bottom-up models consisting only of metric and a single bulk scalar field, 
the scalar potential (\ref{eq:scalarpotential}) controls all the properties of the dual field theory. For example, the quartic coefficient $-1/3$ can be shown to correspond to attractive interactions in elastic two-to-two scattering events \cite{Hoyos:2019kzt,Hoyos:2020fjx} and this coefficient also plays a pivotal role in determining the stiffness. In fact, the value $-1/3$ is precisely the border line case for vanishing trace anomaly and where the equation of state changes from being soft $\partial p/\partial e < {1}/{3}$ to stiff $\partial p/\partial e>{1}/{3}$ \cite{Hoyos:2016cob,Ecker:2017fyh}. Here we therefore also expect that the equation of state could be stiff if the contributions from the rest of the terms in the scalar potential are negligible and we are in the low temperature regime where the bulk scalar has its most important effect. This effect is partially evidenced in the left plot of Fig.~\ref{fig:pair} for larger values for $\phi_Q$.  It would be interesting to generalise our study by relaxing fixed quartic coefficient and explore how the stiffness affects the signal-to-noise ratios in the gravitational wave searches.

More generally, the gravitational wave power spectrum contains information about the equation of state of the underlying theory, in this case a strongly coupled one.  In order to understand how to access this information, one needs to perform 
numerical simulations with the correct equation of state and field-fluid coupling. Up to now, these have been performed only with simplified models of weakly-coupled field theories \cite{Hindmarsh:2017gnf,Hindmarsh:2015qta,Hindmarsh:2013xza}. With holography, one has the exciting prospect of computing the required functions for strongly-coupled theories as well, which will eventually allow gravitational wave detectors to probe the equation of state, and perhaps provide evidence for a 
phase transition in a strongly-coupled theory. We hope to turn to these computations in future work.

\vspace{1.0cm}
\begin{acknowledgments}
We thank Daniel Cutting, Chloe Gowling, Oscar Henriksson, and Javier Tarr\'io for useful discussions. 
M.~H. (ORCID ID 0000-0002-9307-437X) acknowledges support from  the Academy of Finland project number 333609. The work of C.~H.~has been partially supported by the Spanish Ministerio de Ciencia, Innovaci\'on  y Universidades and by the Principado de Asturias through the grants PGC2018-096894-B-100 and GRUPIN-IDI/2018 /000174. N.~J. has been supported in part by the Academy of Finland grant no. 1322307.
\end{acknowledgments}

\appendix

\section{Holographic renormalisation}\label{app:holoreno}

Due to the close nature of the relationship between quantum field theory and gravity through the holographic principle, UV divergences on the field theory side from composite operators approaching coincident points in $d$-dimensions appear as IR divergences on the gravitational side as infinite volumes of $AdS_{d+1}$ geometries. These divergences must be regularised and renormalised, and so the supergravity fields are expanded near the boundary and counterterms must be introduced so as to subtract any divergences that arise. Holographic renormalisation of this type is well known (see e.g.~\cite{bianchi}) and it's the procedure we will follow. The initial step is to expand the solutions at the boundary, and for this the Fefferman-Graham metric
\begin{equation}
ds^2 = \frac{L^2}{u^2}\left(g_{ab}\hspace{0.05cm}\text{d}x^a\text{d}x^b+\text{d}u^2\right)
\end{equation}
is used with the boundary  
at $u\rightarrow0$. Expanding our fields and metric around this point gives
\begin{equation}\label{eq:gexp}
g_{ab} = \gamma_{ab} + g_{ab}^{(2)}u^2+g_{ab}^{(4)}u^4 + \ldots 
\end{equation}
for the boundary metric, and
\begin{equation}
\phi = \Lambda u +\Upsilon u^3 + \ldots 
\end{equation}
for the scalar field. For these variables $\Lambda$ has dimension 1 and $\Upsilon$ has dimension 3, causing $\phi$ to be dimensionless. The leading behaviour of the boundary metric is simply flat Minkowski ($\gamma_{ab}=\eta_{ab}$), and $g_{ab}^{(2)}$ is determined in terms of this and $\Lambda$ as
\begin{equation}
g_{ab}^{(2)} = -\frac{1}{3}\Lambda^2\gamma_{ab}.
\end{equation} 
The $u^4$ coefficient $g_{ab}^{(4)}$ leads to the energy-momentum tensor. Through the holographic dictionary we can also identify $\Lambda$ as the coupling to the dual field operator, and $\Upsilon$ as dual to the vacuum expectation value of the $\Delta$ = 3 operator.
Boundary behaviour in hand, our next job is to tame the divergences by regularising the theory. Denoting the regularised action as $S_{\text{reg}}$, we find the field theory operators through 
\begin{equation}\label{ops}
\langle\mathcal{O}\rangle = \frac{\delta S_{\text{reg}}}{\delta\phi}, \qquad \langle T_{ab}\rangle = \frac{\delta S_{\text{reg}}}{\delta \gamma^{ab}}
\end{equation}
where this action is built up of the Einstein-Hilbert with scalar term, Gibbons-Hawking term, and counterterm as
\begin{equation}
S_{\text{reg}} = S_{EH}+S_{GH}+S_{ct}.
\end{equation}
To find this action the extrinsic curvature is needed for the Gibbons-Hawking term, and in these coordinates that is given by
\begin{equation}
K = -\frac{1}{\sqrt{g_{uu}}}\frac{L^4}{u^4}\partial_{u} \log\left(\frac{L^4}{u^4}\sqrt{-g}\right)\Bigg|_{u\to0} = -\frac{L^3}{u^3}\partial_{u} \log\left(\frac{L^4}{u^4}\sqrt{-g}\right)\Bigg|_{u\to0}.
\end{equation}
We choose the counterterm to regularise the action as a term similar the superpotential, which is usual \cite{Bianchi_2001}. Our counterterm therefore is
\begin{equation}
S_{ct} = \frac{2}{\kappa_5^2}\int d^4x\sqrt{-\gamma}\frac{L^3}{u^4}\left(-\frac{3}{2}-\frac{\phi^2}{2}-\frac{\phi^4}{4\phi_M^2}\right)\Bigg|_{u\to0},
\end{equation}
which excludes the $\phi_Q$ term as this will vanish anyway with $(\phi^6/u^4)|_{u\to0}$ and so would not contribute to the divergences. By choosing this renormalisation scheme we are in effect ``preserving supersymmetry" which causes the vacuum energy to vanish. Combining all together and performing the calculations (\ref{ops}) generates 
\begin{equation}
\langle\mathcal{O}\rangle = -\frac{2L^3}{\kappa_5^2}\left(2\Upsilon+\frac{\Lambda^3}{\phi_M^2}\right) \ ,
\end{equation}
and
\begin{equation}\label{eq:tyymyynyy}
\langle T_{ab}\rangle = \frac{2L^3}{\kappa_5^2}\left\{g_{ab}^{(4)}+\gamma_{ab}\left(\Lambda\Upsilon-\frac{\Lambda^4}{18}+\frac{\Lambda^4}{4\phi_M^2}\right)\right\} \ .
\end{equation}
The tensor which gives rise to the stress-energy tensor, $g_{ab}^{(4)}$, is found to be given by the expression

\begin{equation}
\begin{aligned}
g_{ab}^{(4)} = \textrm{diag}\Big(  - & \frac{1}{36}\left(2\Lambda^4-18\Lambda\Upsilon+27h_4\right),\frac{1}{36}\left(2\Lambda^4-18\Lambda\Upsilon-9h_4\right),  \\
& \frac{1}{36}\left(2\Lambda^4-18\Lambda\Upsilon-9h_4\right),\frac{1}{36}\left(2\Lambda^4-18\Lambda\Upsilon-9h_4\right)\Big) \ ,
\end{aligned} 
\end{equation}  
where $h_4$ is the constant associated with the subleading term in the expansion of the blackening factor $h$ in the limit $\phi\to0$. Recalling that $\gamma_{ab}$ is simply flat Minkowski, the trace of the stress-energy tensor is therefore 
\begin{equation}
\langle T_{a}^{a}\rangle = \frac{2L^3}{\kappa_5^2}\left\{2\Lambda\Upsilon+\frac{\Lambda^4}{\phi_M^2}\right\} \ ,
\end{equation}
where the constant $\Lambda^4$ term has dropped out as well as $h_4$. It is quickly evident that the Ward identity is satisfied by this, giving the form
\begin{equation}
\langle T_{a}^{a}\rangle +\Lambda\langle\mathcal{O}\rangle = 0 \ .
\end{equation}

\section{Numerical methods}\label{app:numerics}

The numerical integration is performed using scipy's ``$\text{solve\_ivp}$" function using the ``Radau'' method, usually with at least 300 points and standard tolerances. It is provided with the system of equations for $G'(\phi)$ and $H'(\phi)$, the Jacobian of the system, and the initial values of $G(\phi_h)$ and $H(\phi_h)$. When numerically integrating, it first must be realised that there is a simple pole at exactly $\phi_h$ in $H'(\phi)$ as $G(\phi_h)=-\frac{4}{3}\gamma_1^h$. Consequently, the solver must be displaced a small amount, $\phi_h+\varepsilon$, by pushing the initial values slightly away from $G(\phi_h)$ and $H(\phi_h)$. The choice of $\varepsilon$ does not seem overly important as long as it is small compared to $\phi_h$ (i.e. $\varepsilon/\phi_h\lesssim1\times10^{-2}-1\times10^{-3}$) but large enough to push the solver away from the pole. The ``$\text{solve\_ivp}$" function is repeated for the desired number of $\phi_h$ values, with careful consideration being taken for the direction of arrays as the system must be integrated from $\phi_h$ down to the minimum value. 

Once solved, the thermodynamic quantities $s$ and $T$ quickly follow from numerical integration (using scipy's ``trapz" function), 
for each value of $\phi_h$. 
Other thermodynamic quantities in this paper are easily derivable from $T$ and $s$. 
Numerical integration and differentiation are subject to errors at large $T$ spacings. 


The critical temperature $T_c$ is determined from the self-intersection of the free energy curve in the $(f,T)$ plane. 
Knowing where $T_c$ is located allows the free energy and energy density curves to be broken up into different branches and found for the symmetric and broken regions, as seen in Fig.~\ref{fig:branches}, enabling calculation of $\alpha$ and $L(T_c)$.
\begin{figure}[h]
  \centering
      \includegraphics[width=0.75\textwidth]{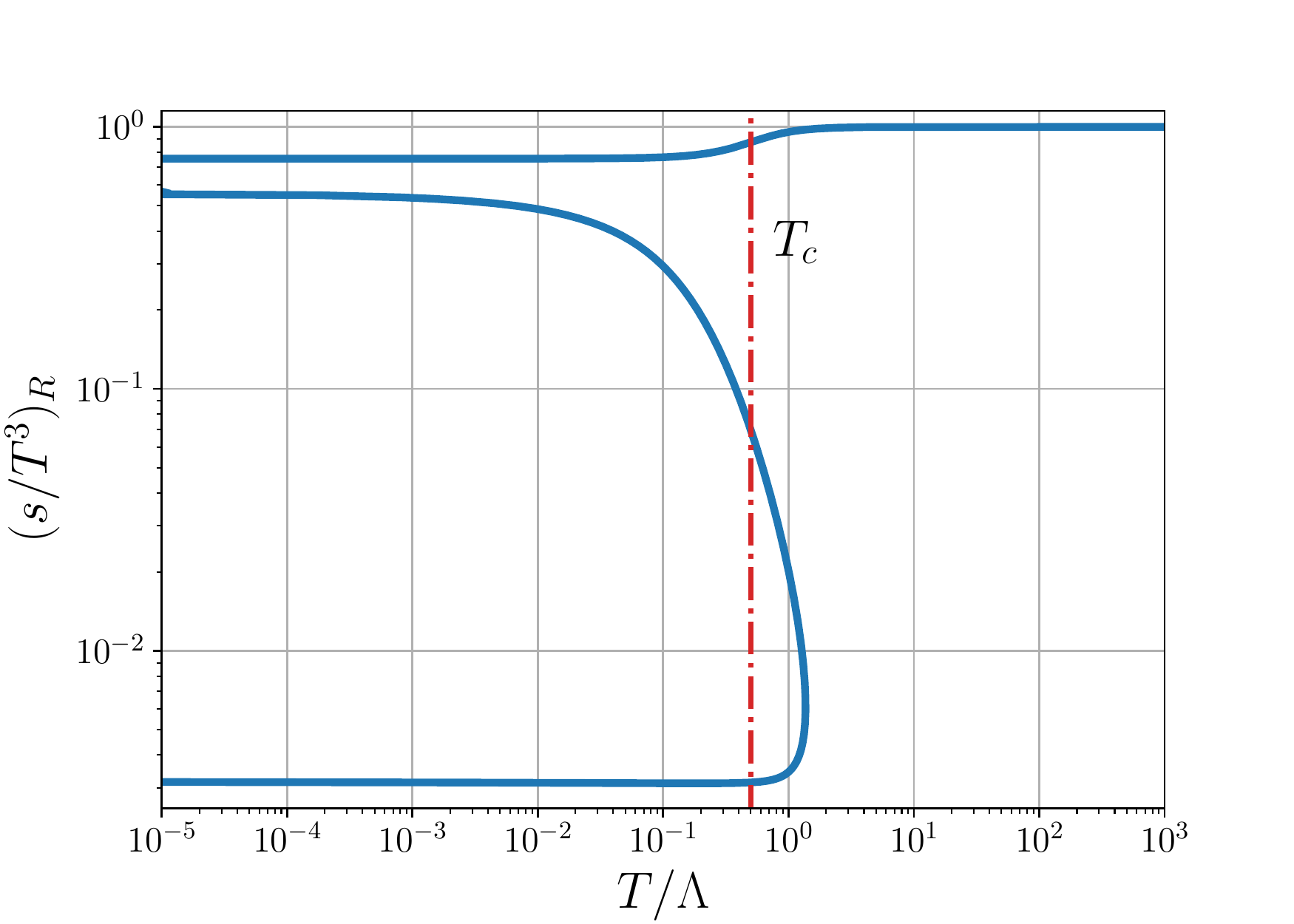}
  \caption{Rescaled entropy versus temperature graph for $\phi_M\simeq0.5797$ and $\phi_Q=10.0$, with critical temperature $T_c/\Lambda \approx0.499$.}
  \label{fig:ent}
\end{figure}

To ensure our numerics performed well, we checked our results against most graphs in \cite{Bea:2018whf}. Fig.~\ref{fig:ent} serves as a comparison with figure~6 in that paper to demonstrate that our numerical solutions are behaving as expected.

\section{Gravitational wave power spectrum model}\label{app:omegagw}

\newcommand{\OmGWnow}{\Omega_{\text{gw},0}}
\newcommand{\OmGWscaled}{\tilde\Omega_\text{gw}}
\newcommand{\fitfun}{C}
\newcommand{\fpnow}{f_{\text{p},0}} 
\newcommand{\HN}{H_\text{n}} 
\newcommand{\Rbc}{R_*} 

The model for the gravitational wave power spectrum from phase transitions 
has undergone a number of changes as understanding has improved. 
We may start by considering the form given in the erratum in Ref. \cite{Hindmarsh:2017gnf}, 
\begin{equation}
\label{eq:corrected}
  \OmGWnow(f) = 2.061 F_{\text{gw},0} K^2 (\HN\Rbc)
\OmGWscaled \fitfun\left(\frac{f}{\fpnow}\right).
\end{equation}
%
%
The form applies only when the sound wave lifetime
$\tau_{\mathrm{sw}} = \Rbc /\sqrt{K}$ is much longer than the Hubble time $\HN^{-1}$.
Ref.~\cite{guo2020phase} showed that modelling the decay as an abrupt switching off at $\tau_{\mathrm{sw}}$ 
introduces a factor 
\begin{equation}
\Upsilon_{\mathrm{sw}}(x) = 1 - \frac{1}{\sqrt{1+2x}} \ .
\end{equation}
where $x = \tau_{\mathrm{sw}}H_N$. 
We take the attenuation timescale to be $\tau_{\mathrm{sw}} = R_*/\sqrt{K}$. 
The LISA Cosmology Working Group (LCWG) model takes the function to be 
$\Upsilon_{\mathrm{sw}}(x) = \min(1,x)$. A similar approximation is seen in \cite{Ellis:2018mja,Ellis_2019,Ellis_2020} with further discussion.
The lack of understanding of how the power spectrum attenuates and changes in form 
is the major uncertainty in this model.

\begin{figure}[h]
 \centering
     \includegraphics[width=1\textwidth]{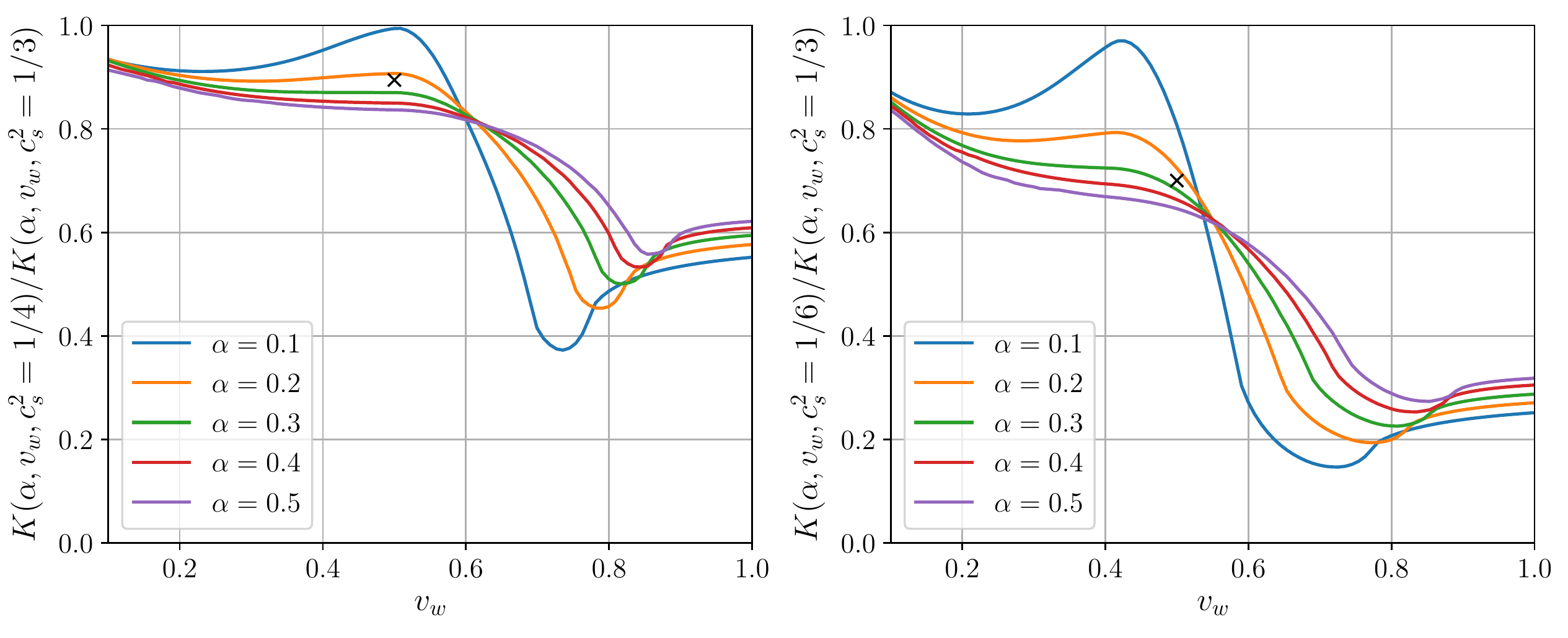}
  \caption{Plots of the effect of varying stiffness of the equation of state $\partial p/\partial e$ on the kinetic energy fraction $K$ for $\alpha = 0.1, 0.2, 0.3, 0.4,$ and $0.5$ as a function of the wall speed $v_w$. This is displayed as a ratio of the kinetic energy fraction for two different equations of state ($\partial p/\partial e =1/4,1/6$) to the kinetic energy fraction of the conformal stiffness $\partial p/\partial e$. Black crosses have been placed on each graph to show a reference point with $v_w=0.5$ and $\alpha=0.25$.}
  \label{fig:c_seff}
\end{figure}

The kinetic energy fraction is estimated from an efficiency factor $\kappa$, 
computed for the self-similar flow around a single expanding bubble of the stable phase \cite{Kamionkowski_1994,Espinosa_2010}. 
It is related to the kinetic energy fraction by
\begin{equation}
K = \frac{\kappa\alpha}{1+\alpha} .
\end{equation}
The LCWG model uses the fitting formulae for $\kappa$ in Ref.~\cite{Espinosa_2010}, which 
are derived in a model with stiffness $\partial p/\partial e = 1/3$.
As we find significant departures from $1/3$, we use instead the 
code snippet given in \cite{giese2020modelindependent}, denoting the values obtained as $K_\text{GKSV}$
Fig.~\ref{fig:c_seff} shows how the kinetic energy fraction is modified as a function of $\alpha$ and $v_w$ 
for stiffnesses $1/4$ and $1/6$. 
%
%
A reference value of $v_w = 0.5$ with $\alpha = 0.25$ has been indicated, for which 
the maximum reduction in the kinetic energy fraction is about $20\%$.

%

\begin{figure}[h]
 \centering
     \includegraphics[width=0.7\textwidth]{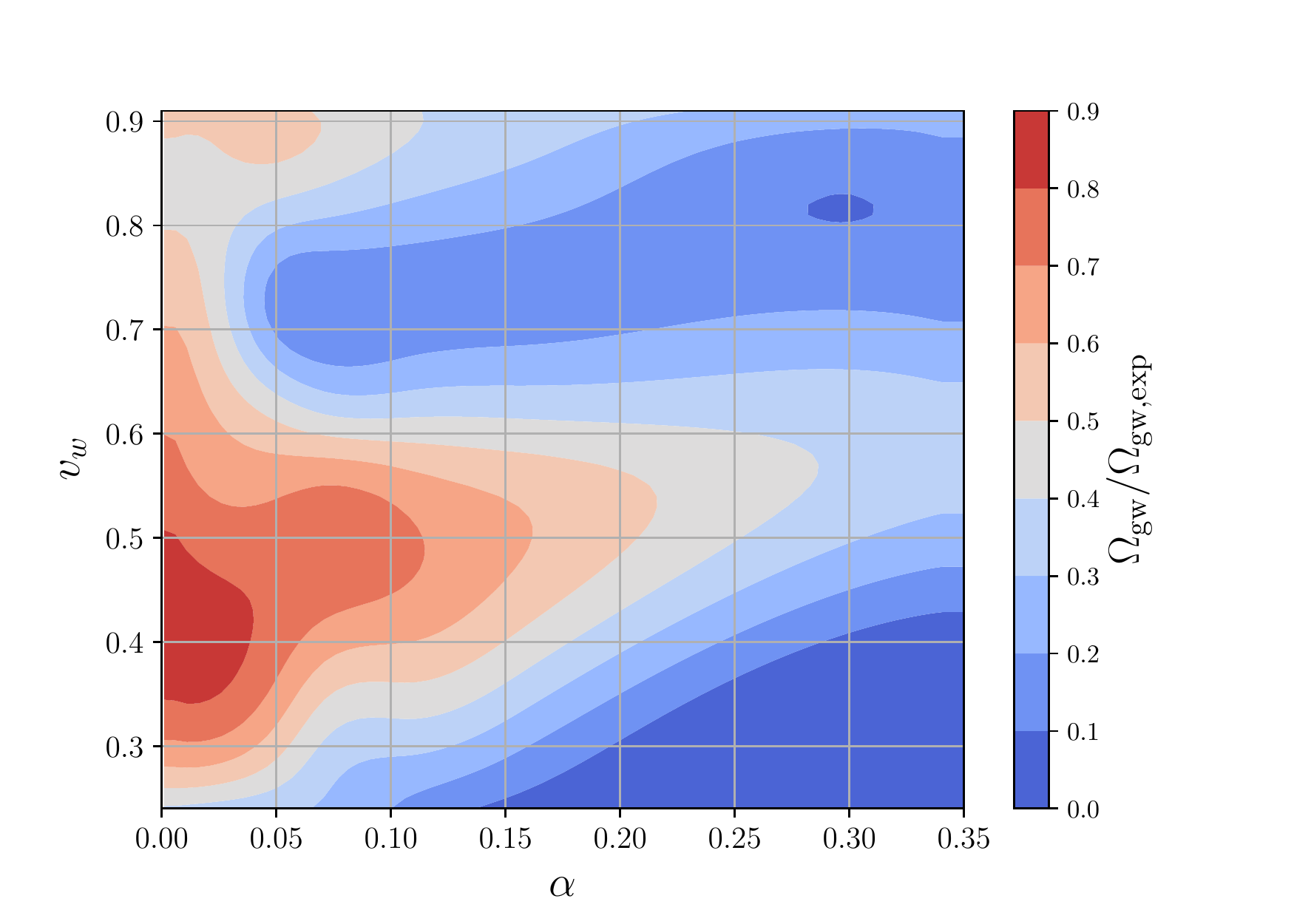}
  \caption{Contours of suppression to $\Omega_{\text{gw}}$ in the form of the ratio $\Omega_{\text{gw}}$ to $\Omega_{\text{gw,exp}}$, where $\Omega_{\text{gw,exp}}$ is the expected power computed according to the 
  LCWG model.}
  \label{fig:damp}
\end{figure}

Finally, 3-dimensional hydrodynamic simulations of strong first-order thermal phase transitions in \cite{Cutting_2020} showed a deficit in kinetic energy compared to the LCWG value \cite{Espinosa_2010}, 
ascribed to slowing of the phase boundary due to reheating of the metastable phase. 
This will have the effect of reducing the gravitational wave signal, and so must be taken into consideration. 
We define a suppression function $\Sigma(v_w,\alpha)$, defined as 
\begin{equation}
\Sigma(v_w,\alpha) = \Omega_{\text{gw}}/\Omega_{\text{gw,exp}},
\end{equation}
where $\Omega_{\text{gw}}$ is the true total gravitational wave power, 
and $\Omega_{\text{gw,exp}}$ is that predicted by the LCWG model. 
We take the values of this function by cubic interpolation of the 
ratio of the last two quantities in Table 1 of \cite{Cutting_2020}. 
Contours of the suppression  function are shown in Fig.~\ref{fig:damp}. 

The final expression for the gravitational wave power spectrum is 
\begin{equation}
  \OmGWnow(f) = 2.061 F_{\text{gw},0} K_\text{GKSV}^2 (\HN\Rbc) 
  \OmGWscaled \fitfun\left(\frac{f}{\fpnow}\right)
  \Upsilon_{\mathrm{sw}}(x) \Sigma(v_w,\alpha) \ .
\end{equation}

\bibliographystyle{JHEP}
\bibliography{Bibfile}
\end{document}